\documentclass[useAMS]{mn2e}
\usepackage{slashbox}
\usepackage{amssymb,amsmath,psfig,times}
\usepackage{graphicx}
\usepackage{graphicx}
\def\gsim{ \lower .75ex \hbox{$\sim$} \llap{\raise .27ex \hbox{$>$}} }
\def\lsim{ \lower .75ex\hbox{$\sim$} \llap{\raise .27ex \hbox{$<$}} }

\def\sc{Schwarzschild}
\def\beq{\begin{equation}}
\def\eeq{\end{equation}}

\def\sc{Schwarzschild}

\def\sw{{\it Swift}}
\def\fe{{\it Fermi}}
\def\gro{{\it CGRO}}
\def\ba{BATSE}

\def\ep{$E_{\rm peak}$}

\def\epo{$E^{\rm obs}_{\rm peak}$}

\def\liso{$L_{\rm iso}$}
\def\eiso{$E_{\rm iso}$}
\def\ama{$E_{\rm peak}-E_{\rm iso}$}
\def\yone{$E_{\rm peak}-L_{\rm iso}$}

\def\th{$\theta_{\rm jet}$}

\title[High-z GRBs]
{Accessing the population of high redshift Gamma Ray Bursts}
\author[Ghirlanda et al.]
{G. Ghirlanda$^{1}$\thanks{E--mail:giancarlo.ghirlanda@brera.inaf.it}, R. Salvaterra$^{2}$,
G. Ghisellini$^{1}$, S. Mereghetti$^{2}$, G. Tagliaferri$^{1}$, S. Campana$^{1}$, \newauthor
J. P. Osborne$^{3}$, P. O'Brien$^{3}$, N. Tanvir$^{3}$, R. Willingale$^{3}$, L. Amati$^{4}$, S. Basa$^{5}$, \newauthor 
M.G. Bernardini$^{1}$, D. Burlon$^{6}$, S. Covino$^{1}$, P. D'Avanzo$^{1}$, F. Frontera$^{7,4}$, D. G\"otz$^{8}$,  \newauthor
A. Melandri$^{1}$, L. Nava$^{9}$, L. Piro$^{10}$, S. D. Vergani$^{11,1}$ \\
$^{1}$INAF -- Osservatorio Astronomico di Brera, via E. Bianchi 46, I-23807 Merate, Italy\\
$^{2}$INAF -- IASF Milano, via E. Bassini 15, I-20133 Milano, Italy \\
$^{3}$Department of Physics and Astronomy, University of Leicester, Leicester
LE1 7RH, UK\\
$^{4}$INAF -- IASF Bologna, via P. Gobetti 101, I-40129 Bologna, Italy\\
$^{5}$ Aix Marseille UniversitŽ, CNRS, LAM (Laboratoire dÕAstrophysique de Marseille) UMR 7326, 13388 Marseille, France\\
$^{6}$The University of Sydney, 44-70 Rosehill Str., Redfern, NSW, 2016\\
$^{7}$Dipartimento di Fisica, Universita' di Ferrara, via Saragat 1, I-44100 Ferrara, Italy\\
$^{8}$CEA Saclay - Irfu/Service d'Astrophysique (AIM), Orme des Merisiers, F-91191, Gif-sur-Yvette Cedex, France\\
$^{9}$Racah Institute of Physics, The Hebrew University of Jerusalem, Jerusalem 91904, Israel\\
$^{10}$INAF -- Istituto Astrofisica e Planetologia Spaziali, Via del Fosso Cavaliere 100, I-00133 Roma, Italy\\
$^{11}$GEPI-Observatoire de Paris Meudon. 5 Place Jules Jannsen, F-92195, Meudon, France
}
\begin{document}

\date{}


\maketitle

\label{firstpage}

\begin{abstract}
Gamma Ray Bursts (GRBs) are a powerful probe of the high redshift Universe. 
We present a tool to estimate the detection rate of high--$z$ GRBs by a generic detector 
with defined energy band and sensitivity. 
We base this on a population model that reproduces the observed properties of 
GRBs detected by \sw, \fe\ and \gro\ in the hard X--ray and $\gamma$--ray bands. 
We provide the expected cumulative distributions of the flux and fluence 
of simulated GRBs in different energy bands. 
We show that scintillator detectors, operating at relatively high energies (e.g. tens 
of keV to the MeV), can detect only the most luminous GRBs at high redshifts due to the 
link between the peak spectral energy and the luminosity (\yone) of GRBs. 
We show that the best strategy for catching the largest number of high--$z$ 
bursts is to go softer (e.g. in the soft X--ray band) but with a very high sensitivity. 
For instance, an imaging soft X--ray detector operating in the 0.2--5 keV energy band 
reaching a sensitivity, corresponding to a fluence, of $\sim10^{-8}$ erg cm$^{-2}$ 
is expected to detect $\approx$ 40 GRBs yr$^{-1}$ sr$^{-1}$ at $z\ge 5$ 
($\approx$ 3 GRBs yr$^{-1}$ sr$^{-1}$ at $z\ge 10$). 
Once high--$z$ GRBs are detected the principal issue is to secure their redshift.  
To this aim we estimate their NIR afterglow flux at relatively early times and evaluate 
the effectiveness of following them up and construct usable samples 
of events with any forthcoming GRB mission dedicated to explore the high--$z$ 
Universe. 
\end{abstract}


\section{Introduction}

The study of the Universe before and during the epoch of reionization represents one of the 
major themes for the next generation of space and ground--based observational facilities. 
Many questions about the first phases of structure formation in the early Universe are still open: 
{\it When and how did first stars/galaxies form? What are their properties? 
When and how fast was the Universe enriched with metals? How did reionization proceed?}

Thanks to their brightness, Gamma Ray Bursts (GRBs) can be detected up to very high redshift, as already 
shown by the detection of GRB~090423 at $z=8.2$ (Salvaterra et al. 2009; Tanvir et al. 2009) 
and GRB~090429 at $z\sim 9.4$ (Cucchiara et al. 2011). 
Thus, GRBs represent a complementary, and to some extent unique, tool to study the early 
Universe (see e.g. McQuinn et al. 2009; Amati et al. 2013). 
A statistical sample of high--$z$ GRBs can provide fundamental information about: 
the number density and properties of low-mass galaxies (Tanvir et al. 2012; Basa et al. 2012; 
Salvaterra et al. 2013; Berger et al. 2014); the neutral--hydrogen fraction 
(Gallerani et al. 2008; Nagamine et al. 2008; McQuinn et al 2008; Robertson \& Ellis 2012); 
the escape fraction of UV photons from high--$z$ galaxies (Chen et al. 2007; Fynbo et al. 2009); 
the early cosmic magnetic fields (Takahashi et al. 2011), and the level of the local 
inter--galactic radiation field (Inoue et al. 2010). 
Moreover, GRBs will also allow us to measure independently the cosmic star--formation rate, 
even above the limits of current and future galaxy surveys (Ishida et al. 2011), provided 
that we are able to quantify the possible biases present in the observed GRB redshift 
distribution (Salvaterra et al. 2012; Trenti et al. 2013; Vergani et al. 2014). 
Even the detection of a single GRB at $z>10$ would be extremely useful to constrain dark 
matter models (de Souza et al. 2013) and primordial non--Gaussianities (Maio et al. 2012). 
Finally, it has been argued that the first, metal--free stars (the so--called PopIII stars) 
can result in powerful GRBs (Meszaros \& Rees 2010; Komisarov \& Barkov 2010; Suwa \& Yoka 2011; 
Piro et al. 2014). 
Thus GRBs offer a powerful route to directly identify such elusive objects and study the 
galaxies in which they are hosted (Campisi et al. 2011). 
Even indirectly, the role of PopIII stars in enriching the first galaxies with metals can 
be studied by looking to the absorption features of PopII GRBs blowing out in a medium 
enriched by the first PopIII supernovae (Wang et al. 2012; Ma et al. 2014 in prep).

Specific predictions on the detection rate of high--$z$ GRBs (Bromm \& Loeb 2002; Gorosabel et al. 2004; Daigne, Rossi \& Mochkovitch 2006; 
Salvaterra \& Chincarini 2007; Salvaterra et al. 2008; Butler et al. 2010; Qin et al. 2010; 
Salvaterra et al. 2012; Littlejohns et al. 2013) are of fundamental importance for the design of future missions. 
In particular, it is often believed that the population of high redshift GRBs could be easily 
accessed with soft X--ray instruments, due to the $(1+z)^{-1}$ cosmological redshift of their characteristic 
peak spectral energy. 
However, the peak energy is strongly correlated with the burst luminosity/energy (Yonetoku et al. 2004; Amati et al. 2002) and, 
therefore, the trigger energy band and the detector sensitivity are strongly interlaced. 

\begin{center}
\begin{figure}
\includegraphics[scale=0.45,trim=1cm 8cm 0.7cm 8cm, clip]{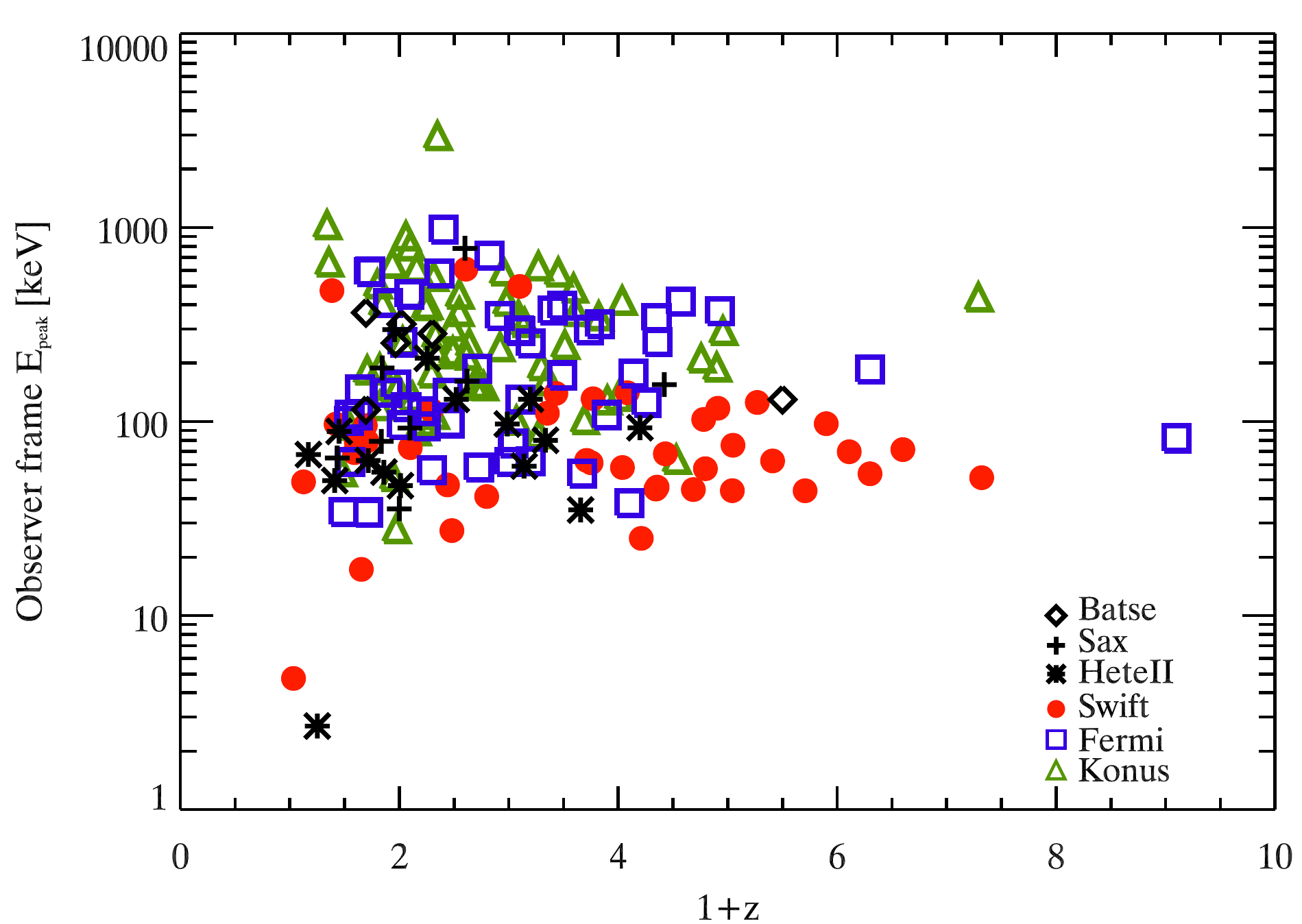}
\caption{Observer frame peak energy of the $\nu F_{\nu}$ spectrum versus redshift of 185 GRBs with 
measured redshift $z$ and well determined prompt emission \ep. 
Different symbols (as displayed in the legend) correspond to detectors/instruments that provided the measure of \ep.}
\label{fg0}
\end{figure}
\end{center}

The typical prompt emission peak energy \epo\ of GRBs detected by past/present instruments extends 
from a few keV up to a few MeV. 
The distribution of the observer frame peak energy versus redshift of the 185 GRBs with measured spectral parameters and redshift $z$ (up to September 2014) is shown in Fig.~\ref{fg0} 
(see also Gruber et al. 2011). 
The  spectroscopically confirmed  highest redshift GRB 090423 ($z=8.2$) has an observer frame \epo$\sim$90 keV (von Klienin et al. 2009) 
as measured by \fe--GBM. 
Fig.~\ref{fg0} shows the absence of GRBs with low measured \epo\  at high redshifts.
This is a consequence of the spectral--luminosity correlation \yone\ (where \liso\ is the isotropic 
equivalent peak luminosity -- Yonetoku et al. 2004) and the limited sensitivity of  current GRB detectors. 
In other words, we can access, at high redshifts, only  to the most luminous GRBs that have the highest \ep. 
This compensates the $(1+z)^{-1}$  decrease of \ep\ expected if \ep\ were independent of luminosity. 

This paper aims to study how these two instrumental properties can be optimized for detecting 
the largest number of high redshift bursts. 
We base that on a population synthesis code (\S2) anchored to the observed properties 
(prompt emission flux and fluence and optical afterglow flux) of the GRBs detected by \sw, \fe, \ba\ (\S3). 
In \S4 we provide a general tool (\S4)  
that allows to predict the detection rate of GRBs for a detector with specified energy band and sensitivity. 
Beside their detection, we also study the possibility to follow up the high--$z$ 
GRB population in the X--ray and NIR bands through the characterization of their afterglow flux level (\S4). 
A standard cosmology with $h=\Omega_{\Lambda}=0.7$ and $\Omega_{\rm M}=0.3$ is assumed. 

\section{Simulation setup} 
The main assumptions to simulate the burst population are:
\begin{enumerate}
\renewcommand{\theenumi}{(\arabic{enumi})} 

\item GRBs are distributed in redshift (up to $z=20$) following the 
GRB formation rate:
\begin{equation}
\psi(z)\propto (1+z)^{\delta} \psi_{\star}(z)
\label{z1}
\end{equation}
where $\psi_{\star}(z)$ is the comoving cosmic star formation rate 
(Li 2008; see also Hopkins \& Beacom 2008):
\begin{equation}
\psi_{\star}(z)=\frac{0.0157+0.118z}{1+(z/3.23)^{4.66}}
\label{z}
\end{equation}
where $\psi(z)$ has units of ${\rm M}_{\sun}$ ${\rm yr}^{-1}$  ${\rm Mpc}^{-3}$. 
Salvaterra et al. (2012 -- S12 hereafter) have shown that in order to reproduce the observed 
$z$ distribution of the \sw\ complete sample and the \ba\ count distribution $\delta=1.7$ is 
needed. We adopted this value. This model does not include the possible contribution at $z\ge10$ 
of GRBs produced from popIII stars.  
The density evolution of the GRB formation rate is supported by two considerations: (a) it is  consistent with the metallicity bias which appears from recent studies of the GRB hosts (e.g. Vergani et al. 2014), and (b) it is consistent with the expectation of the collapsar model.

\item The assumed GRB luminosity function (LF) is a broken power law (as derived in S12):
\begin{eqnarray}
\phi(L_{\rm iso})&\propto&\left(L_{\rm iso}\over L_{\rm cut}\right)^{a}; \quad
                            L_{\rm iso}\le L_{\rm cut} \nonumber\\
	             &\propto& \left(L_{\rm iso}\over L_{\rm cut}\right)^{b}; \quad
	                        L_{\rm iso}> L_{\rm cut}
\label{lf}
\end{eqnarray}
where the parameter values $[a,b,L_{\rm cut}]=[-1.50, -2.32, 3.8\times 10^{52} 
{\rm erg\, s^{-1}}]$ were constrained by S12, through the BAT6 sample. 
The LF is assumed to extend between minimum and maximum luminosities 
$10^{49}$ erg s$^{-1}$ and $10^{55}$ erg s$^{-1}$ (Pescalli et al. 2014). 
The results for the rate of high--$z$ GRBs are almost insensitive to the choice of 
these bracketing values since the LF is steeply decaying at high luminosities and the 
low luminosity limit affects only the dim bursts that are most likely unseen by any conceivable detectors. 

An alternative model (e.g. Salvaterra et al. 2009; 2012) assumes that what evolves with redshift is the break 
$L_{\rm cut}$ of the luminosity function rather than the GRB formation efficiency. We have tested also 
this assumption and show in \S5 that our conclusions are unchanged.  

\item To compute the flux and fluence of GRBs in a given energy range, 
we need to associate a spectrum to each simulated burst. 
Assuming that all bursts have a spectrum described by a double power law function 
(Band et al. 1993) smoothly joined at the peak energy $E_{\rm peak}$, we assign to 
each simulated burst a low energy photon spectral index $\alpha$ and a high energy photon 
spectral index $\beta$ drawn from Gaussian distributions centered at $-1$ and $-2.3$ with 
$\sigma=0.2$, respectively (e.g. Kaneko et al. 2006; Nava et al. 2011; Goldstein et al. 2012). 

\item The  spectral peak energy $E_{\rm peak}$ is obtained through the \yone\ correlation (Yonetoku et al. 2004). 
\liso\ represents the peak isotropic luminosity, i.e. related to the flux at the peak of the $\gamma$--ray light curve. 
We adopt the \yone\ correlation obtained with the complete BAT6 sample (Nava et al. 2012):
$\log{(E_{\rm peak})} = -25.33 + 0.53 \log{(L_{\rm iso})}$. 
We account for the scatter $\sigma=0.29$ dex of the data points around this correlation (Nava et al. 2012). 

\item Similarly, in order to have also a fluence associated to each simulated burst, 
we adopt the \ama\ correlation (Amati et al. 2002) between the peak energy and the 
isotropic equivalent energy $E_{\rm iso}$, as derived with the BAT6 
sample (Nava et al. 2012): 
$\log{(E_{\rm peak})} = -29.6 + 0.61 \log{(E_{\rm iso})}$. 
Also in this case we consider the scatter $\sigma=0.25$ dex.

Note that our simulation approach does not need to specify the distribution of \ep\ from where assign the values of this parameter to the simulated bursts. Indeed, the starting assumption of the simulation is the luminosity function. Therefore, the simulated GRB population will have a distribution of \liso\ which is the form of the assumed luminosity function. It is the assumed \yone\ correlation which transforms the \liso\ distribution into the distribution of \ep. 

The latter passage is not a 1:1 mapping since we allow for the scatter of the correlations. 
For both the \ama\ and the \yone\ correlation the scatter that we adopt represents the $\sigma$ of the 
gaussian distribution of the distance of the data points measured perpendicular to the best fit line of the correlations.  
It is still debated if and how much these correlations are affected by selection effects (Band \& Preece 2005; Nakar \& Piran 2005; 
Butler et al. 2007; Butler, Kocevski \& Bloom 2009; Shahmoradi \& Nemiroff 2011; Kocevski 2012) despite some tests suggest
that this might not be the case (Bosnjak et al. 2008; Ghirlanda et al. 2008; Nava et al. 2008; Amati, Frontera \& Guidorzi 2009; Krimm et al. 2009; Ghirlanda et al. 2012; Dainotti et al. 2014; Mochkovitch \& Nava 2014). However, there seems to be an agreement that, if not properly a correlation, there could be a boundary in the \ama\ and \yone\ plane which divides the plane in two regions: the right hand side of the correlations lacking of bursts with large \eiso\ and \liso\ (whose lack might be due to some physical reason -  e.g. Nakar \& Piran 2005) and the left hand side of the correlations  (corresponding to GRBs with low \eiso\ and \liso) which might be affected by instrumental biases. If so, the plane could be uniformly filled with bursts on the left hand side of the correlations (i.e. at intermediate/low \eiso, \liso)  with no correlation between \eiso\ and \ep\ and \liso\ and \ep. A fraction of these events could be missed due to some instrumental selection effects that shapes its cut in the plane so to make an apparent correlation. We have tested also this hypothesis and discuss it in \S5.

\end{enumerate} 

With these assumptions we derive the 1--s peak flux $P$ of each simulated burst in a 
given energy range $\Delta E=[E_{1},E_{2}]$: 
\begin{equation}
P_{\Delta E}=\frac{L_{\rm iso}}{4\pi d_{L}(z)^{2}} 
\cdot \frac{\int_{E_{1}}^{E_{2}} N(E)dE}{\int_{1{\rm keV}/(1+z)}^{10^4{\rm keV}/(1+z)} E\, N(E)dE}
\label{flux}
\end{equation}
where $N(E)$ is the observer frame photon spectrum and $d_{L}(z)$ is the luminosity distance 
corresponding to the redshift $z$. 
The fluence $F$ is: 
\begin{equation}
F_{\Delta E}=\frac{E_{\rm iso}(1+z)}{4\pi d_{L}(z)^{2}} \cdot \frac{
 \int_{E_{1}}^{E_{2}} E\, N(E)dE}{\int_{1{\rm keV}/(1+z)}^{10^4{\rm keV}/(1+z)} E\, N(E)dE}
\label{fluence}
\end{equation}

The integrals at the denominators are performed over the rest frame 1keV--10MeV energy range, which is typically adopted to derive \eiso\ and \liso.

\section{Simulation normalization and consistency checks}
\begin{figure*}
\hskip -0.4truecm
\includegraphics[scale=0.5,trim=2cm 8cm 2cm 4cm, clip]{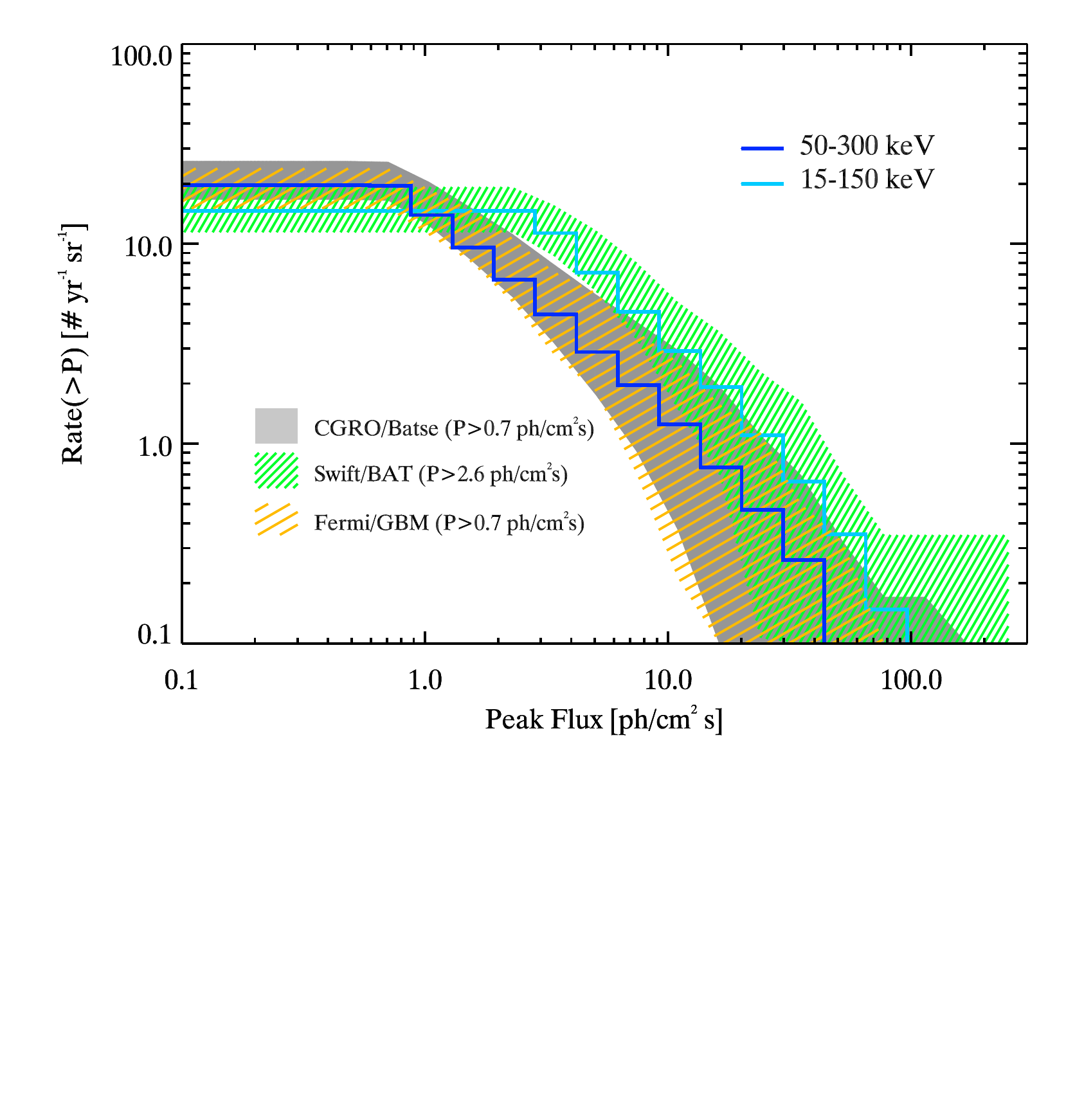}
\includegraphics[scale=0.5,trim=2cm 8cm 2cm 4cm, clip]{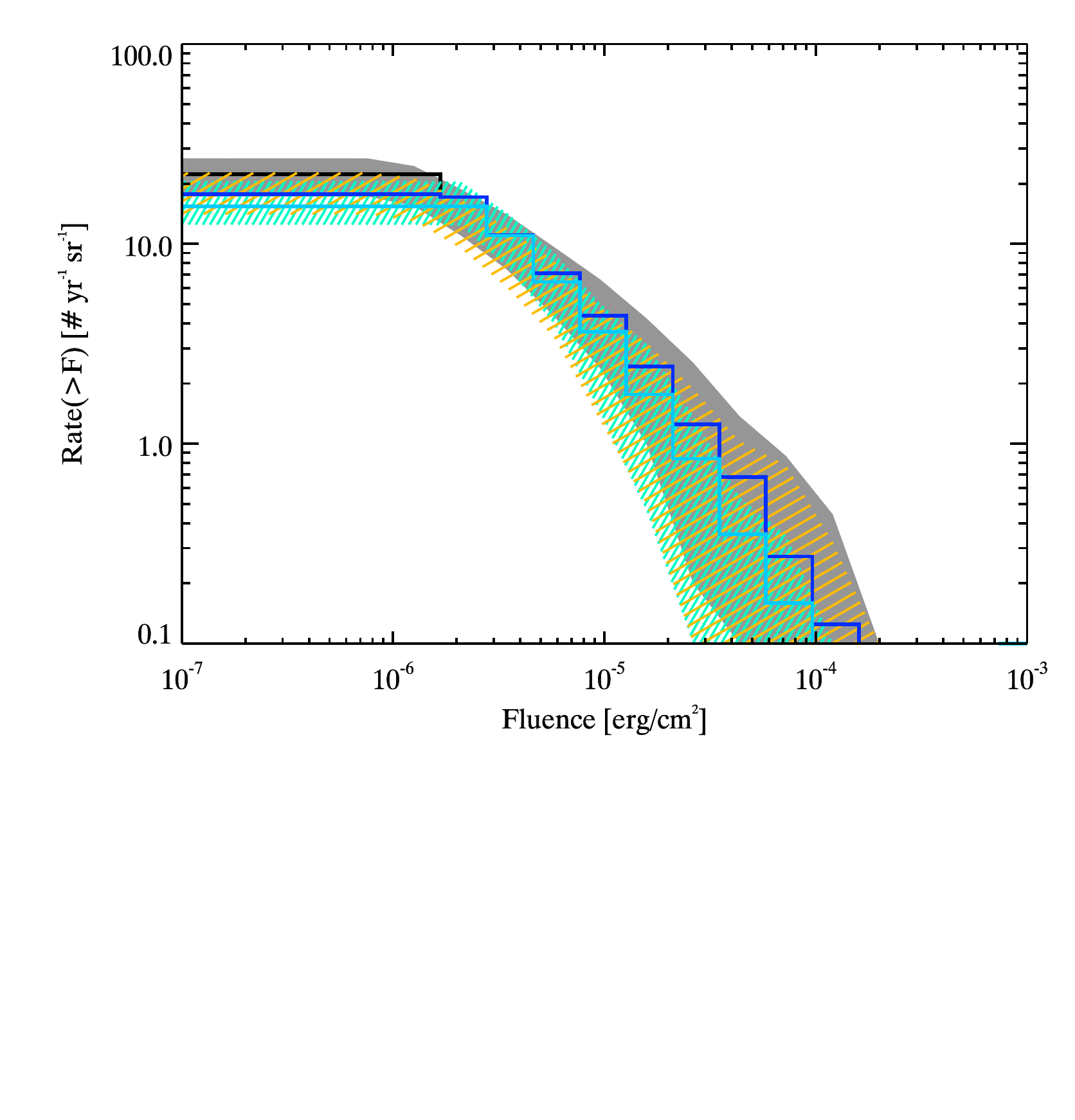}
\vskip -3truecm
\includegraphics[scale=0.5,trim=2.5cm 9cm 2cm 4cm, clip]{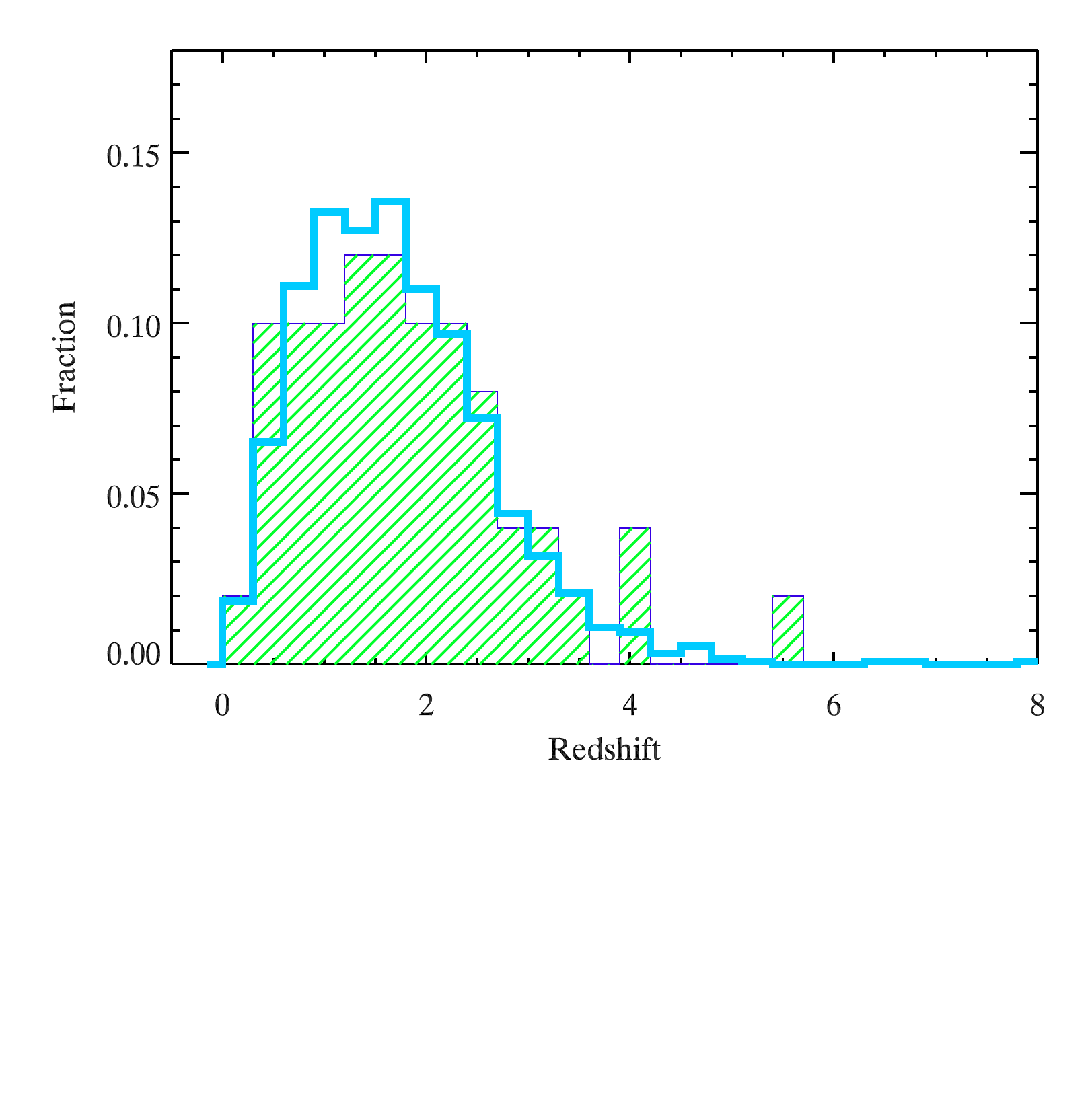}
\includegraphics[scale=0.5,trim=2.5cm 9cm 2cm 4cm, clip]{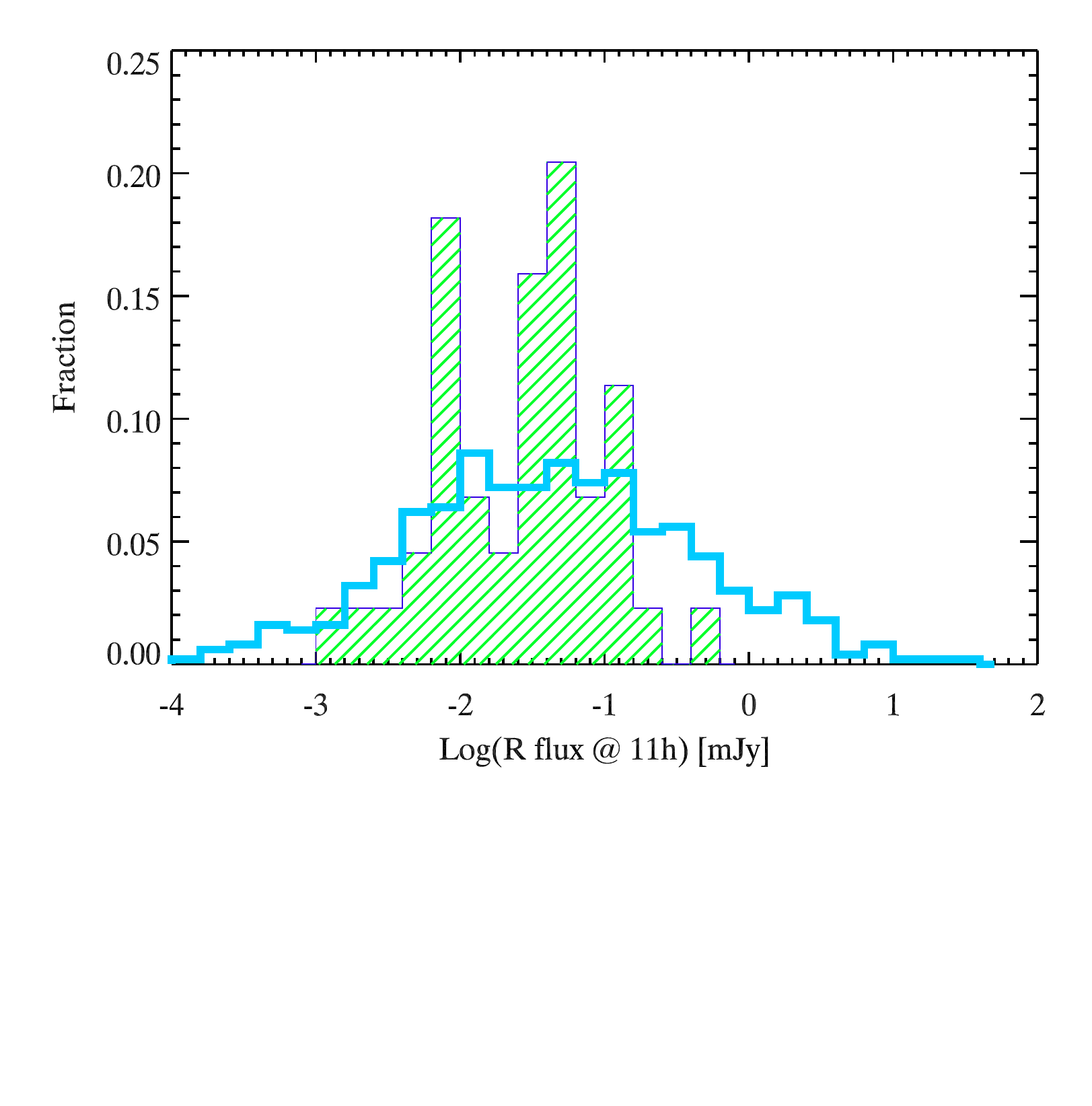}
\vskip -2.5truecm
\caption{
Comparison of the simulated GRB population (solid lines) with real samples of bursts 
(solid filled and hatched histograms). 
{\it Top left panel:} 
cumulative flux distribution of the bright \sw\ bursts (hatched green region -- one $\sigma$ 
contingency of the distribution), of the \gro--\ba\ 4B catalog (solid filled grey region) 
and of the \fe-GBM 2--year catalog (hatched orange region). 
\sw\ fluxes are integrated in the the 15--150 keV range, while  \ba\ and GBM fluxes in the 
50--300 keV energy range. 
The blue, black and cyan solid lines show the flux distribution  (in the corresponding energy range) 
of the simulated GRB population corresponding to \fe, \ba\ and \sw. 
{\it Top right panel:} cumulative fluence distributions. 
Same color coding and symbols of  the left panel. 
{\it Bottom left panel:} redshift distribution of the BAT6 sample (hatched histogram) and 
of the simulated sample with the same flux limit $P\ge$2.6 ph cm$^{-2}$ s$^{-1}$ (solid cyan line). 
{\it Bottom right panel:} extinction corrected optical afterglow flux (in the $R$ band) at 11h of the BAT6 sample 
(from Melandri et al. 2014) compared to the simulated population with the same peak flux cut (solid cyan line).  
}
\label{fg1}
\end{figure*}
\begin{figure*}
\hskip 0.8truecm
\includegraphics[scale=0.9,trim=10 350 10 80,clip=true]{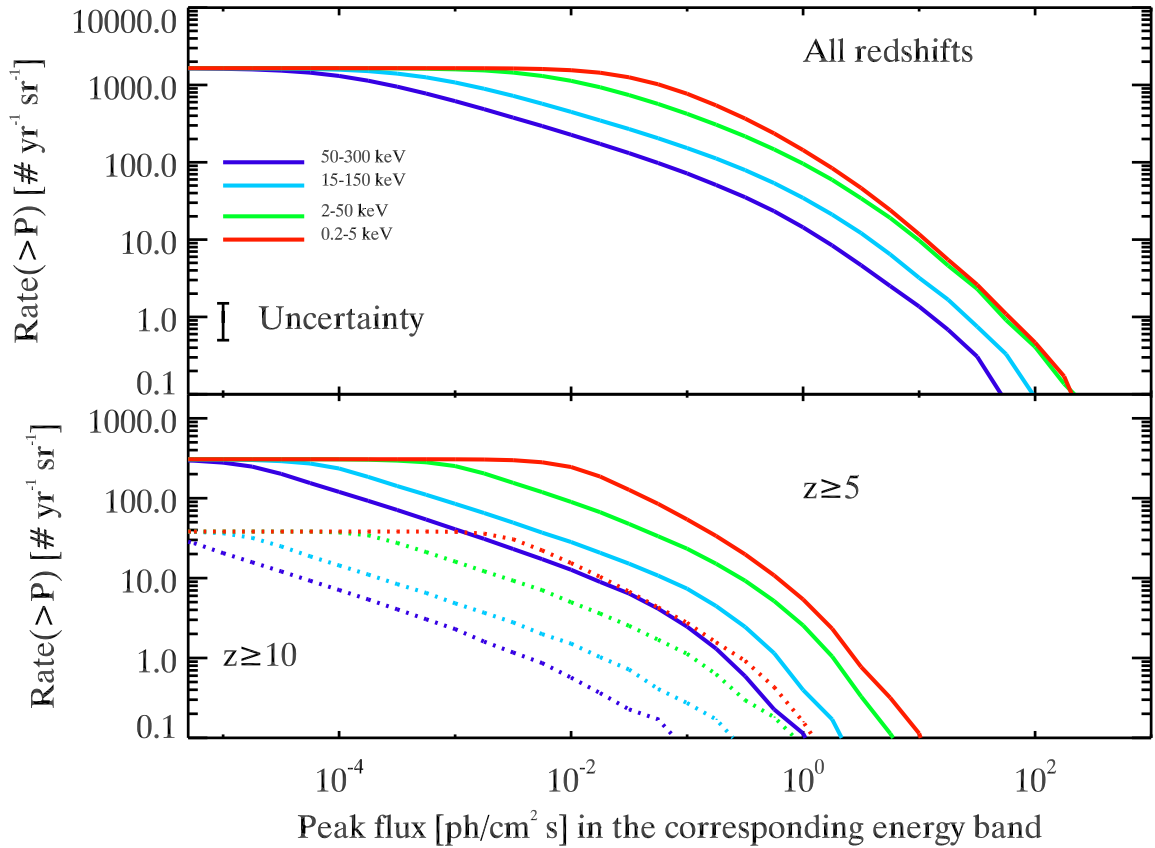}
\vskip 0.5truecm
\caption{
Cumulative peak flux distributions representing the rate, i.e. the number of 
GRBs yr$^{-1}$ sr$^{_2}$, with peak flux $\ge P$. 
Different energy ranges where the peak flux $P$ is computed are shown with different colors (as shown in the legend).  
The typical uncertainty on the predicted rates (for any energy range) is shown by the vertical bar (see \S5).
{\it Top panel}: entire population of simulated GRBs. 
{\it Bottom panel}: population of high redshift GRBs: $z\ge5 (10)$ shown by the solid (dotted) line. 
For the soft energy band 0.2--5 keV (shown by the red line) a typical $N_{H}=2\times10^{21}$ cm$^{-2}$ 
has been assumed to account for the absorption.
}
\label{fg2}
\end{figure*}

\begin{figure*}
\hskip 0.8truecm
\includegraphics[scale=0.9,trim=10 350 10 80,clip=true]{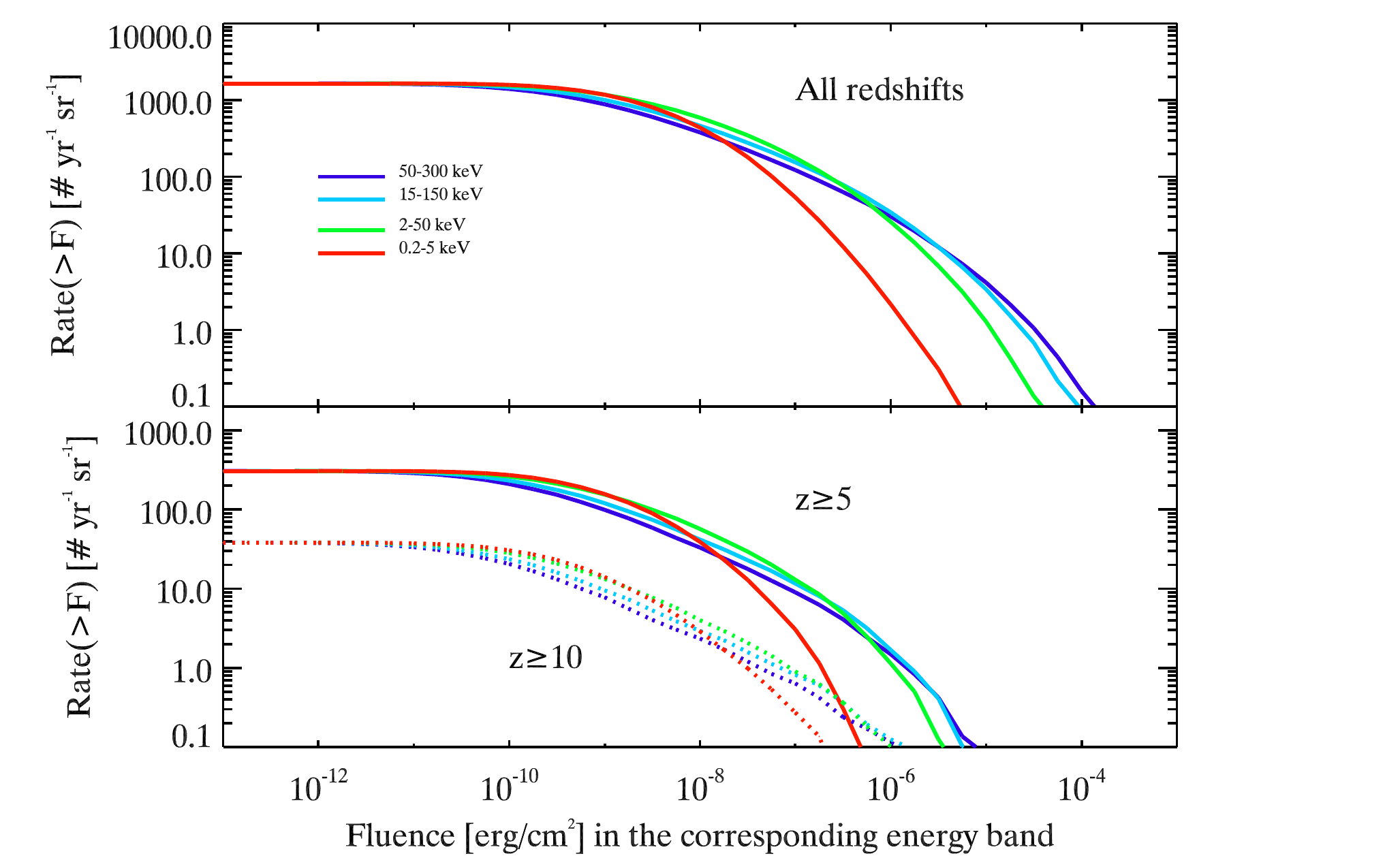}
\vskip 0.9truecm
\caption{
Cumulative fluence distributions representing the rate, i.e. the number 
of GRBs yr$^{-1}$ sr$^{-1}$, with fluence $\ge F$. 
Different energy ranges where the fluence $F$ is computed are shown with different 
colors (as shown in the legend). 
{\it Top panel}: entire population of simulated GRBs. 
{\it Bottom panel}: population of high redshift GRBs: $z\ge5 (10)$ shown by the solid (dotted) line. 
For the soft energy band 0.2-5 keV (shown by the red line) a typical $N_{H}=2\times10^{21}$ cm$^{-2}$ 
has been assumed to account for the absorption. 
}
\label{fg3}
\end{figure*}

The simulated population of GRBs is normalized to the actual population of bright bursts detected by \sw/BAT. 
The BAT6 sample of S12 considered GRBs detected by \sw/BAT with peak flux $P\ge 2.6$ ph cm$^{-2}$ s$^{-1}$ 
(integrated in the 15--150 keV energy range) selected only for their favorable observing 
conditions (Jakobsson et al. 2006). 
The sample contains 58 GRBs and is 95\% complete in redshift. 

We have extracted from the current \sw\ sample (773 events detected until July 2014 
with $T_{90}\ge2$ s)\footnote{http://swift.gsfc.nasa.gov/docs/swift/archive/grb\_table/} 
the 204 GRBs with 1--s peak flux $P \ge 2.6$ ph cm$^{-2}$ s$^{-1}$ (i.e. the same flux 
threshold used to define the BAT6). 
Considering a typical field of view of 1.4 sr and a mission lifetime of approximately 9.5 years, 
we estimate a detection rate by \sw\ of $\sim$15 events  yr$^{-1}$ sr$^{-1}$ 
with peak flux $P\ge$2.6 ph cm$^{-2}$ s$^{-1}$. 
We normalize the simulated GRB population to this rate.  

The peak flux cumulative rate distribution of the \emph{real} \sw\ sample of 204 GRBs 
with $P \ge 2.6$ ph cm$^{-2}$ s$^{-1}$ is shown in the top panel of Fig. \ref{fg1} (green hatched region representing the 
statistical uncertainty on the rate). \sw\ has detected many more GRBs with peak flux smaller than this limit which is about six times larger than the actual \sw\ flux limit. The consistency of the simulated population with the entire \sw\ sample is discussed in \S4.1.
The corresponding distribution of the 
\emph{simulated} GRB population is represented by the cyan solid line in Fig.\ref{fg1}.  
As shown in the bottom left panel of Fig.\ref{fg1}, the simulated bursts with  
$P_{\rm [15-150]} \ge 2.6$ ph cm$^{-2}$ s$^{-1}$ (cyan solid line) have a redshift distribution 
consistent with that of the real BAT6 sample (hatched histogram).
 
Note that the comparison of the population of simulated bursts with the real swift GRB sample shown in Fig.2 (top panels and bottom left panel) are consistency checks. Indeed, their scope is to show that the assumptions (\S2), on which the code is based, are correctly implemented and reproduce the population of bursts (i.e. \sw\ bright events) from which they are derived.

\subsection{The \fe\ GBM and \ba\ samples}

An interesting test is to verify if the simulated GRB population is representative 
also of larger samples of GRBs detected by other instruments besides \sw--BAT. 

The Gamma Burst Monitor (GBM) on board \fe\ has  detected many bursts and systematic analysis 
of their spectral properties exist (Nava et al. 2011; Goldstein et al. 2012; Gruber et al. 2014). 
The first two years GBM sample (Goldstein et al. 2012) contains 478 GRBs with the corresponding 
spectral parameters, peak flux $P$ and fluence $F$. 
We considered the flux integrated in the 50--300 keV energy range and cut the GBM sample at 
$P\ge 0.7$ ph cm$^{-2}$ s$^{-1}$ in order to overcome the possible incompleteness of 
the sample at lower fluxes, obtaining 293 GBM bursts. 
The GBM is an all sky monitor that observes on average $\sim$70\% of the sky. 
Therefore, the average GBM detection rate of bursts, with $P\ge 0.7$ ph cm$^{-2}$ s$^{-1}$, 
is 17 events  yr$^{-1}$ sr$^{-1}$. 
Fig. \ref{fg1} (top left panel) shows that the 50--300 keV  
flux distribution of real GBM bursts (hatched orange region) is consistent with that of the
simulated population (blue solid line). 

We have considered also the \ba\ 4B catalog (Meegan et al. 1998) containing 1496 bursts 
with estimated peak flux $P$ and fluence $F$ (both integrated in the 50--300 keV energy range). 
Also in this case we cut the sample to $P\ge 0.7$ ph cm$^{-2}$ s$^{-1}$ (i.e. the same level of 
the GBM), to overcome for incompleteness at low fluxes, and obtain 964 events. 
Considering an average $\sim$75\% portion of the sky observed by \ba, the detection rate 
is $\sim$22 events yr$^{-1}$ sr$^{-1}$  above the considered flux limit. 
The \ba\ peak flux distribution (shaded grey region in Fig. \ref{fg1}) is 
consistent with the simulation (blue solid line in the same figure).

The top right panel of Fig. \ref{fg1} shows the cumulative fluence distributions of the \sw, \ba\ and GBM samples 
as shaded regions (representing the 1$\sigma$ statistical uncertainties). 
The peak flux and fluence distributions of GBM and \ba\ are similar 
(see Nava et al. 2011a)\footnote{The three catalogs contain some bursts with only 
their peak flux or only their fluence reported. 
This is the reason why the cumulative peak flux and fluence distributions saturate at slightly 
different levels for the same instrument.}. The solid lines in the top right panel of Fig. \ref{fg1} represent the 
simulated GRB population. 

\subsection{The afterglow}

The study of high redshift GRBs relies on the capabilities of securing a redshift for these events. 
This depends on the brightness of their afterglow in the NIR band. 
In order to study and predict the afterglow flux of the simulated GRB population, 
we implemented in the code the afterglow emission model as developed in van Eerten (2010, 2011). 
This corresponds to the afterglow emission from the forward shock of a decelerating fireball 
in a constant interstellar density medium. 
We assume a constant efficiency $\eta=0.2$ of conversion of kinetic energy to radiation during 
the prompt phase and a typical jet opening angle of 7$^\circ$ (corresponding to the mean value 
of the reconstructed distribution of \th\ -- Ghirlanda et al. 2013). 
For the ISM density we assume a Gaussian distribution centered at $n=3$ cm$^{-3}$ 
and extending from 1 to 30 cm$^{-3}$. 

We use the afterglow $R$ flux distribution of the BAT6 sample (Melandri et al. 2014 -- corrected 
for extinction, see also Covino et al. 2013) to fix the parameters of the external shock 
micro--physics (e.g. Sironi, Spitkovsky \& Arons 2013):  the fraction of energy that the external shock shares with electrons and 
magnetic field, $\epsilon_{e}$ and $\epsilon_{B}$ and the slope of the shock accelerated electrons $p$. 
Assuming $\epsilon_{e}=0.02$, $\epsilon_{B}=0.008$  and $p=2.3$ (which also reproduce the radio 
afterglow flux of the BAT6 sample - Ghirlanda et al. 2013b) we show that the $R$ flux 
(computed at 11h) distribution of the simulated bright bursts (i.e. with $P\ge 2.6$ 
ph cm$^{-2}$ s$^{-1}$ -- cyan solid line in the bottom right panel of Fig. \ref{fg1}) 
is consistent with that of the BAT6 sample (hatched green histogram). 
A Kolmogorov--Smirnov test results in a probability of 0.05 that the two distributions are 
drawn from the same parent population. 
Differently from the other panels, the aim of the bottom right panel of Fig.\ref{fg1} is that of fixing the free afterglow 
parameters of the simulated GRB population in order to match the observed flux distribution of the BAT6 sample of GRBs 
(solid filled histogram). We do not pretend to explore the afterglow parameter space here (which 
would require detailed modelling of individual bursts - e.g. Zhang et al. 2014) but rather fix the parameters (assumed equal 
for all simulated bursts) in order to predict the typical afterglow flux of the simulated burst population. Therefore, 
we consider the agreement of the simulated (solid line) and real GRB population afterglow flux distribution 
shown in the bottom panel of Fig.\ref{fg1} satisfactory. 
With these parameters we can predict the afterglow emission of the simulated GRB population 
at any time and any frequency.

\section{Results}

GRBs are detected by current instruments as a significant increase of the count rate with respect 
to the average background. 
This is the rate trigger and it is typically implemented by measuring the count rate on several 
timescales and different energy bands (e.g. Lien et al. 2014 for \sw--BAT). 
\sw--BAT also adopts  the image trigger mode: this searches for any new  source in images acquired 
with different integration times. In general both triggers are satisfied by most GRBs 
but the image trigger is particularly sensitive to dim/long GRBs: in the \sw\ GRB sample 
the relatively small fraction of bursts with only image triggers have a smaller peak flux 
than the rest of the bursts (Toma et al. 2011). 
The time dilation effect at particularly high redshifts  makes GRB pulses appear longer. 
Therefore, image trigger can better detect high--$z$ events. 

The rate trigger is particularly sensitive to the ``spikiness" of GRBs and, therefore, to 
the peak flux $P$, while the  image trigger is sensitive to the time integrated flux of the 
burst, i.e. its fluence.  
For these reasons, we present here the peak flux $P$ and fluence $F$ distributions of 
the simulated GRB population.

Both $P$ and $F$ can be computed on a specific energy range $\Delta E$ in our code (Eq.~\ref{flux}, \ref{fluence}). 
To present our results, based on the past/present GRB detectors which 
were mainly either collimated scintillators or imaging coded masks, we have considered  the 50--300 keV 
energy range where both \ba\ and GBM operate (representative of scintillator detectors) and the 15--150 
keV energy range of \sw--BAT (representative of coded masks). 
For the predictions of the detection of high--$z$ GRBs we considered, for reference, the soft energy 
range 2--50 keV (the {\it BeppoSAX} Wide Field Cameras - Jager et al. 1997 - were actually operated in the 2--20 keV energy range  
and similarly the Wide X-ray Monitor - Kawai et al. 1999 - on board {\it Hete--II}), where a  coded mask instrument could operate and the very soft range 0.2--5 keV, appropriate for an instrument like Lobster (Osborne et al. 2013).

\subsection{Peak flux and Fluence}

We have computed the peak flux in the four energy bands given above for the entire population of simulated GRBs. 
The cumulative peak flux distributions are shown in Fig. \ref{fg2}. 
We show the rate of GRB in units of yr$^{-1}$ sr$^{-1}$. 
The total rate (where the curves saturates at low fluxes) is $\sim$ 1600 GRBs  yr$^{-1}$ sr$^{-1}$. 
The flux where the curves saturates depends on the energy range where $P$ is computed. 
The curve showing the flux in the 0.2--5 keV band (red line) is slightly different with respect to the other bands. 
Indeed, to account for the absorption in the soft X--ray band, we have applied a systematic correction of 
the flux by a factor 1.32, which corresponds to the ratio between the intrinsic and absorbed flux for a 
typical power law photon spectrum ($N(E)\propto E^{-\Gamma}$)  with photon index $\Gamma=1$ and assuming 
an average $N_{H}=2\times 10^{21}$ cm$^{-2}$ at $z=0$ (Evans et al. 2009; Willingale et al. 2013). 
The bottom panel of Fig. \ref{fg2} shows the cumulative peak flux distribution for the population of 
GRBs at $z\ge 5$ (solid lines) and at $z \ge 10$ (dotted lines). 

The curves in Fig. \ref{fg2} can be used to predict the detection rate of an instrument 
operating in any of the four energy bands considered once its flux limit is known. 
A note on the use of Fig. \ref{fg2}: since fluxes in different energy ranges are represented, 
the  threshold (i.e. sensitivity limit of a given instrument) should be applied to the corresponding curve. 

For instance, assuming that the flux limit of \sw--BAT is $\sim$0.4 ph cm$^{-2}$ s$^{-1}$ in the 15--150 keV 
energy range, the top panel of Fig. \ref{fg2} shows that (cyan line) this limit corresponds to 
$\sim$56 GRBs  yr$^{-1}$ sr$^{-1}$. 
Considering an average field of view of 1.4 sr and a 9.5 yr mission lifetime, this rate corresponds 
to 745 GRBs detected by \sw\ with $P\ge0.4$ ph cm$^{-2}$ s$^{-1}$ (computed in the 15--150 keV energy band). 
This is in  agreement with the $\sim$730 real GRBs (with $T_{90}\ge 2$ s) present in the \sw\ on line
catalog with peak flux above this limit. 
Similarly one can use the bottom panel of Fig. \ref{fg2} to compute the rate of GRBs at high redshift that 
can trigger \sw.
We predict that the rate of GRBs at $z\ge5$ detectable by \sw\ at a level of  
$P\ge$0.4 ph cm$^{-2}$ s$^{-1}$ in the 15--150 keV should be 1.2 yr$^{-1}$ sr$^{-1}$ 
(corresponding to $\sim$ 16 GRBs over the present mission lifetime consistent with the observed number).

In the \fe\ GRB population, instead, assuming a flux limit of 0.7 ph cm$^{-2}$ s$^{-1}$ in the 50--300 keV 
energy range and considering the corresponding (blue) curve in Fig. \ref{fg2}, we predict a rate 
of 15 yr$^{-1}$ sr$^{-1}$ over all redshifts (top panel of Fig.~\ref{fg1}),
and a rate of 0.1 yr$^{-1}$ sr$^{-1}$ at $z\ge 5$ (bottom panel of Fig.~\ref{fg1}). 
These rates are consistent with the \fe\ detection rate above this flux limit. 

Fig. \ref{fg3} shows the fluence cumulative distribution of the simulated population. 
Color codes are as in Fig. \ref{fg2}. 
We note that the curves corresponding to different energy ranges are very similar at relatively low fluence values. 
This is because we show here the energy fluence, i.e. in units of erg cm$^{-2}$ (as typically 
reported in GRB databases). 
In the appendix (Fig. A1) the same cumulative curves are shown with the photon fluence 
(i.e. in units of ph cm$^{-2}$) which might also be useful in estimating the detection rate of 
GRBs for an instrument working with image triggers.
As explained for the peak flux, each curve represented in Fig. \ref{fg2} should be 
compared with the fluence limit in the corresponding energy range.

\subsection{High energy detector triggers}
\begin{figure}
\includegraphics[scale=0.5,trim=1.5cm 8cm 2cm 9cm, clip]{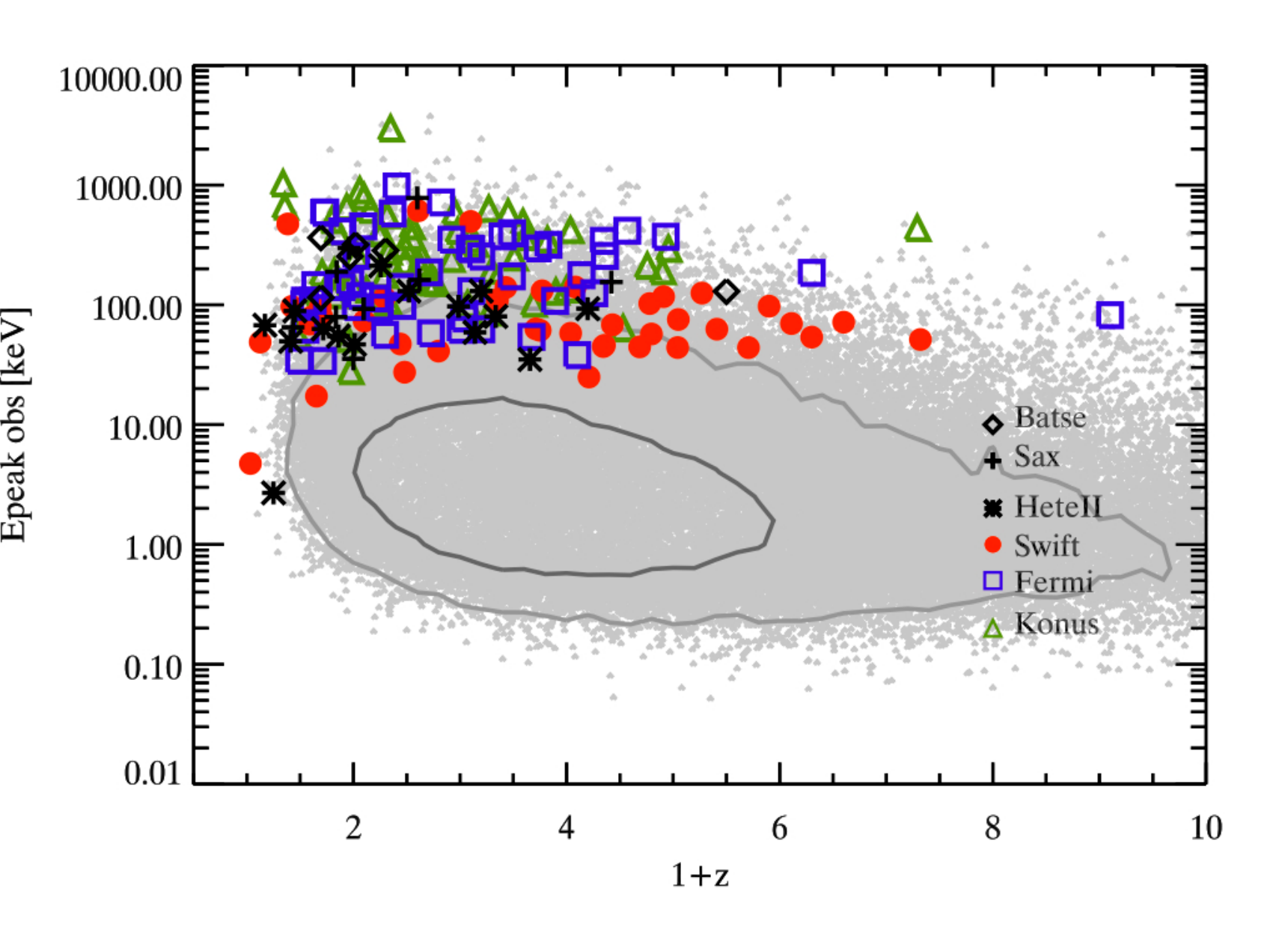}
\caption{
Observer frame peak energy versus redshift. 
Real GRBs with measured $z$ and well constrained \ep\ (185 GRBs) are shown with different symbols 
(as described by the legend). 
The grey cloud of points represents the simulated GRB population.
The dark(light) grey contour represents the 68(95)\% containment region of the simulated bursts. 
The plot is truncated at $z=9$ for clarity. 
}
\label{fg5}
\end{figure}
\begin{figure}
\includegraphics[scale=0.5,trim=1.8cm 8cm 2cm 9cm, clip]{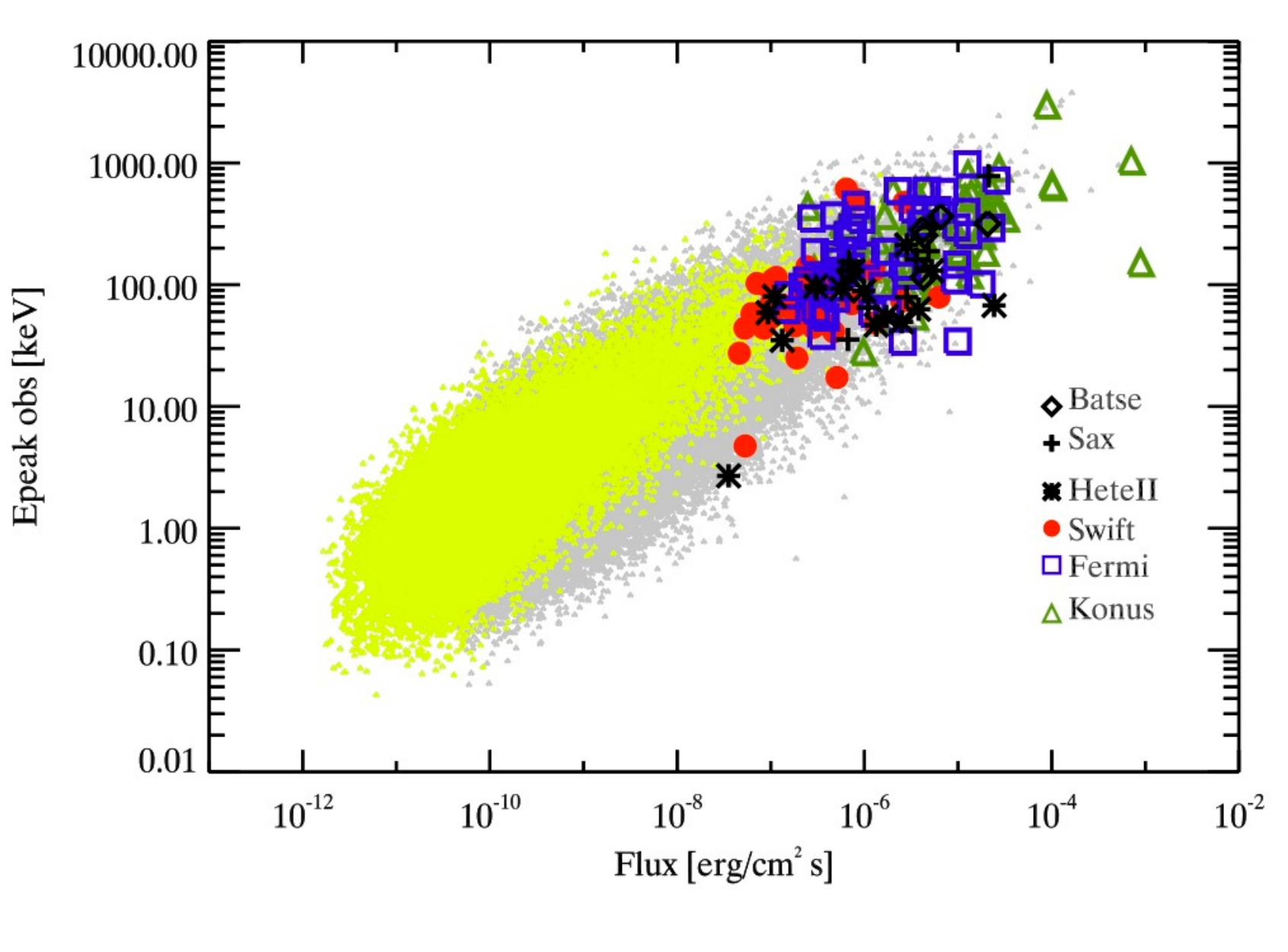}
\caption{
Observer frame peak energy versus bolometric flux. 
Real GRBs with measured $z$ and well constrained prompt emission spectral peak energy 
(185 GRBs) are shown with different symbols.  
The grey cloud of points represents the simulated GRB population. 
Yellow points are GRBs at $z>5$. 
}
\label{fg6}
\end{figure}
\begin{figure}
\hskip-0.2truecm
\includegraphics[scale=0.43,trim=0.5cm 7cm 1cm 8cm, clip]{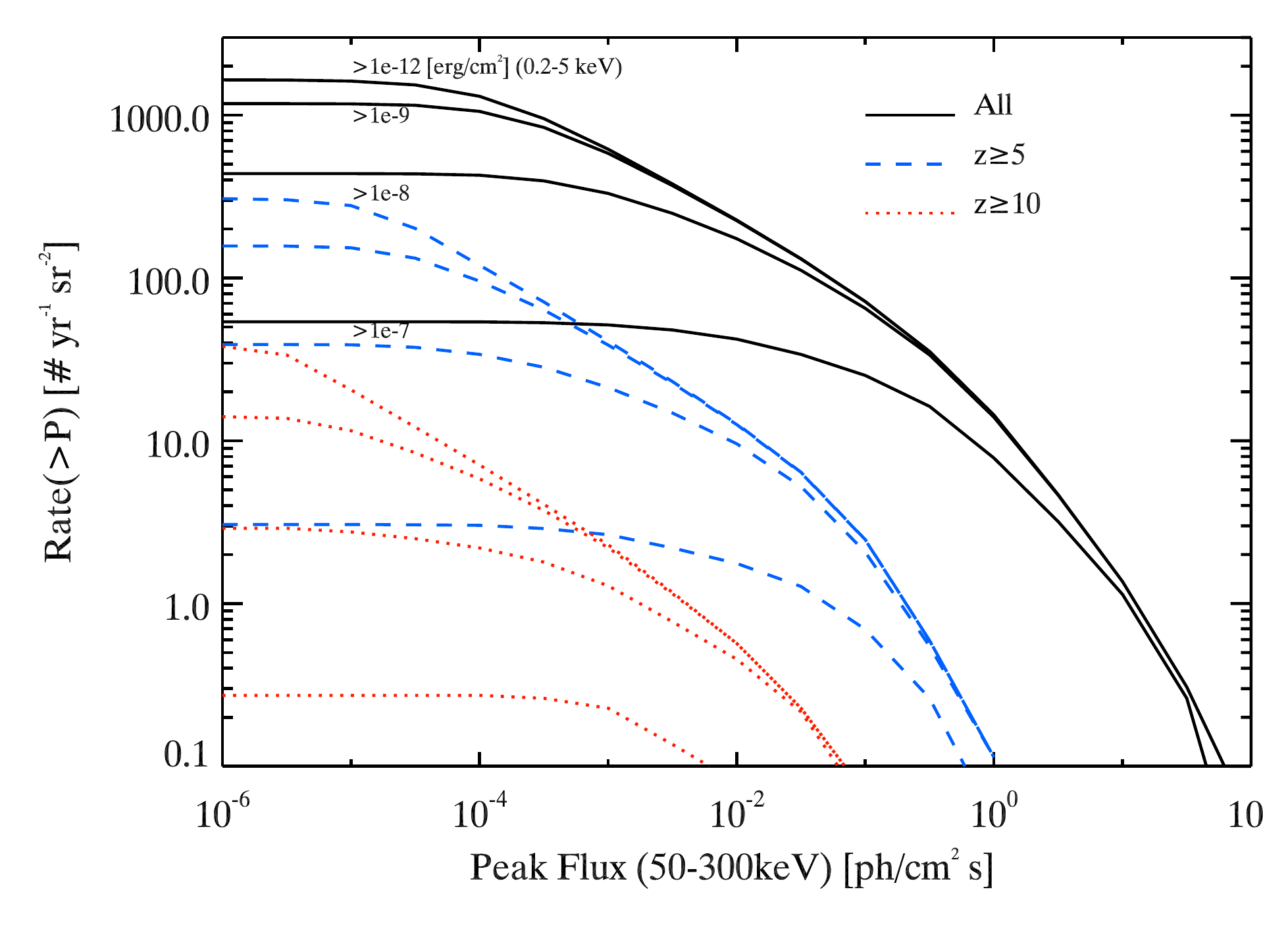}
\caption{
Cumulative peak flux (in the 50--300 keV energy range) distribution for different cuts 
of the soft (0.2--5 keV) energy fluence (as shown for the four solid lines). 
The solid, dashed and dotted lines show the entire, $z\ge5$ and $z\ge10$ population of simulated bursts. 
The four lines for each set correspond to the 0.2--5 keV fluence cuts of $F\ge$ 
$10^{-7}$, $10^{-8}$,  $10^{-9}$ and  $10^{-12}$ erg cm$^{-2}$. 
These curves serve to estimate the rate of common triggers for two instruments: 
one working in the 0.2--5 keV energy range and one in the 50--300 keV energy range, 
once both have assigned fluence and peak flux thresholds, respectively.  
}
\label{fg7}
\end{figure}
 
Current GRB detectors, operating at energies $\ge$10 keV, are detecting the 
most luminous GRBs at high redshifts. 
Indeed, due to the \yone\ correlation, these  events have relatively high peak energies.
Fig. \ref{fg5} shows the observer frame peak energy of real GRBs detected by different instruments 
(as labeled) versus their redshift (same as Fig.~\ref{fg1}) together with the  entire simulated GRB population 
(grey region). 
At high $z$ the detection of the most luminous bursts, due to the \yone\ correlation, 
also selects the GRBs with the largest peak energies. 
This effect can also be seen in Fig. \ref{fg6} which shows the observer frame peak energy versus 
the bolometric flux for the real 185 GRBs with well constrained spectral parameters and 
for the simulated GRB population (same symbols of Fig. \ref{fg5}). 
Here we also highlight (yellow region) the simulated population of GRBs at $z\ge5$. 
We note that, due to the \yone\  correlation, a correlation is present in the observer 
frame between the observed peak energy and the peak flux (e.g. Nava et al. 2008). 
Through Fig. \ref{fg6} one can understand that actual GRB detectors, most sensitive 
from 10 keV up to a few MeV, select those bursts with peak energy in this energy range. 
These events, due to the existence of the \yone\ correlation, are also the brightest 
(and most luminous) bursts although their number is shaped by an overall decreasing luminosity function. 

Fig.~\ref{fg6} shows that in order to efficiently detect high redshift GRBs (yellow symbols) 
one needs a detector working in a softer energy band (e.g. 0.1-10 keV) but also much more 
sensitive than current detectors. 

We have considered four energy bands $\Delta E$ based on two classes of detectors: 
imaging coded mask detectors, typically operated up to few hundreds of keV and 
scintillator detectors operating in 10 keV--1 MeV range. 
It is clear that the hunt for high--$z$ GRBs is optimized with detectors operated in the soft 
energy range, i.e. few keV. 
However, it is worth exploring the rate of detections of GRBs (and high--$z$ bursts) by both types of detectors. 
Indeed, for a configuration with both these detectors, for instance, while the detection rate of 
high--$z$ bursts is maximized by the soft energy detector, the study of the GRB prompt 
emission properties (spectral shape, luminosity etc.) is possible for those bursts 
also detected by the high energy detector. 

Fig. \ref{fg7} shows the 50--300 keV peak flux cumulative rate distribution of the simulated 
GRB population for different cuts of their 0.2--5 keV energy fluence (as labeled). 
We also show the rate for GRBs at $z\ge5$ and $z\ge10$ (blue dashed and red dotted lines, respectively). 
For a soft X--ray detector (working in the 0.2--5 keV energy range) with a sensitivity threshold 
of $10^{-8}$ erg cm$^{-2}$ coupled with a typical scintillator detector (working in the 50--300 
keV energy range) with a sensitivity of  $10^{-1}$ phot cm$^{-2}$ s$^{-1}$, the rate of GRBs 
detected by {\it both} detectors is $\sim$65 yr$^{-1}$ sr$^{-1}$  and  2 yr$^{-1}$ sr$^{-1}$ at $z\ge5$.

\subsection{Afterglow}

\begin{figure*}
\hskip -1truecm
\includegraphics[scale=0.8,trim=0.5cm 7cm 0.5cm 6cm, clip]{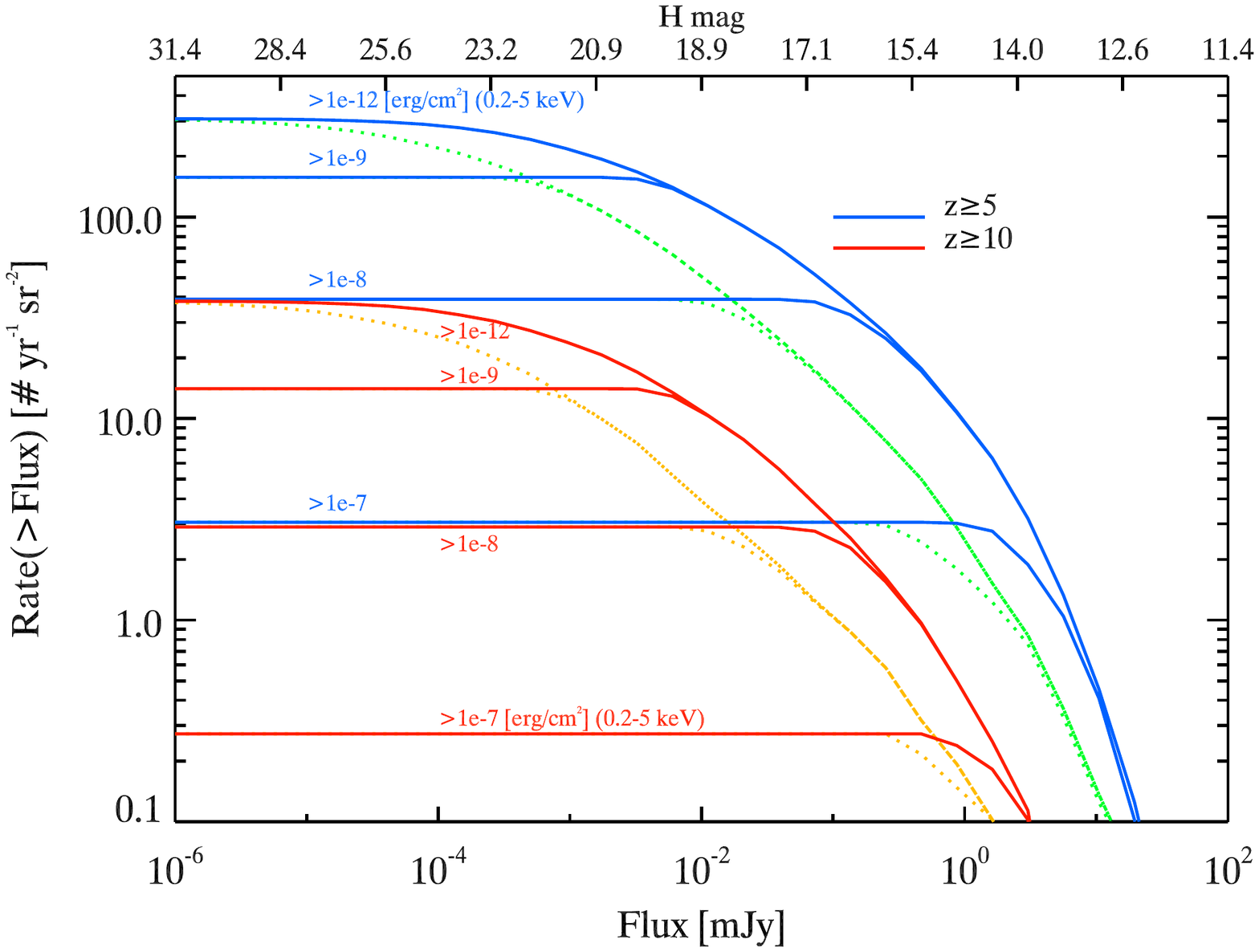}
\caption{
Cumulative NIR  flux distribution of the afterglow of GRBs at $z>5$ and $z>10$ (blue and 
red lines respectively) at 500 s for different cuts on the 0.2--5 keV fluence.  
For  $5<z<10$ we considered the $J$ band and for $z>10$ the $H$ band. 
We  assume $Av=0$ for the population of high redshift GRBs. 
The dotted lines show the $J$ flux at 1 h.  
}
\label{fg4}
\end{figure*}

In addition to the detection of the high redshift GRB population, it is fundamental to assess their redshifts. 
This is possible through early time NIR follow up observations. 
To this aim we have computed the NIR flux of the simulated GRB population as described in \S3. 
Fig. \ref{fg4} shows the cumulative $J$ band flux distribution for the GRBs at $z\ge5$ (solid lines) 
and $z\ge10$ (dashed lines) considering different thresholds of their 0.2--5 keV energy fluence. 

A way to optimize the follow up of high redshift GRBs could be with a dedicated NIR
telescope on board any GRB dedicated satellite trough rapid slewing. Assuming a 1 m telescope in space we expect 
H$\sim$22.5 in 500 seconds of integration and H$\sim$22 in 1h for imaging and low (R=20) resolution 
spectroscopy. Higher resolution spectroscopy (e.g. R$\sim$1000) , needed for metallicity studies, 
can be performed for afterglows brighter than H$\sim$18. With these estimates we evaluate, for imaging and assuming a rapid follow up i.e. 
within 500s,  a rate of $\sim$40 GRBs yr$^{-1}$ sr$^{-1}$ at $z\ge5$ and 3 yr$^{-1}$ sr$^{-1}$ at $z\ge10$ 
(with 0.2--5 keV fluence $\ge 10^{-8}$ erg cm$^{-2}$). Metallicity studies could be performed for $\sim$30(1.5) GRBs yr$^{-1}$ sr$^{-1}$ at $z\ge5$(10) (considering a GRB detector with a limiting 0.2--5 keV fluence $\ge 10^{-8}$ erg cm$^{-2}$).

\begin{figure}
\hskip -1truecm
\includegraphics[scale=0.53,trim=0.5cm 14cm 2cm 4.2cm, clip]{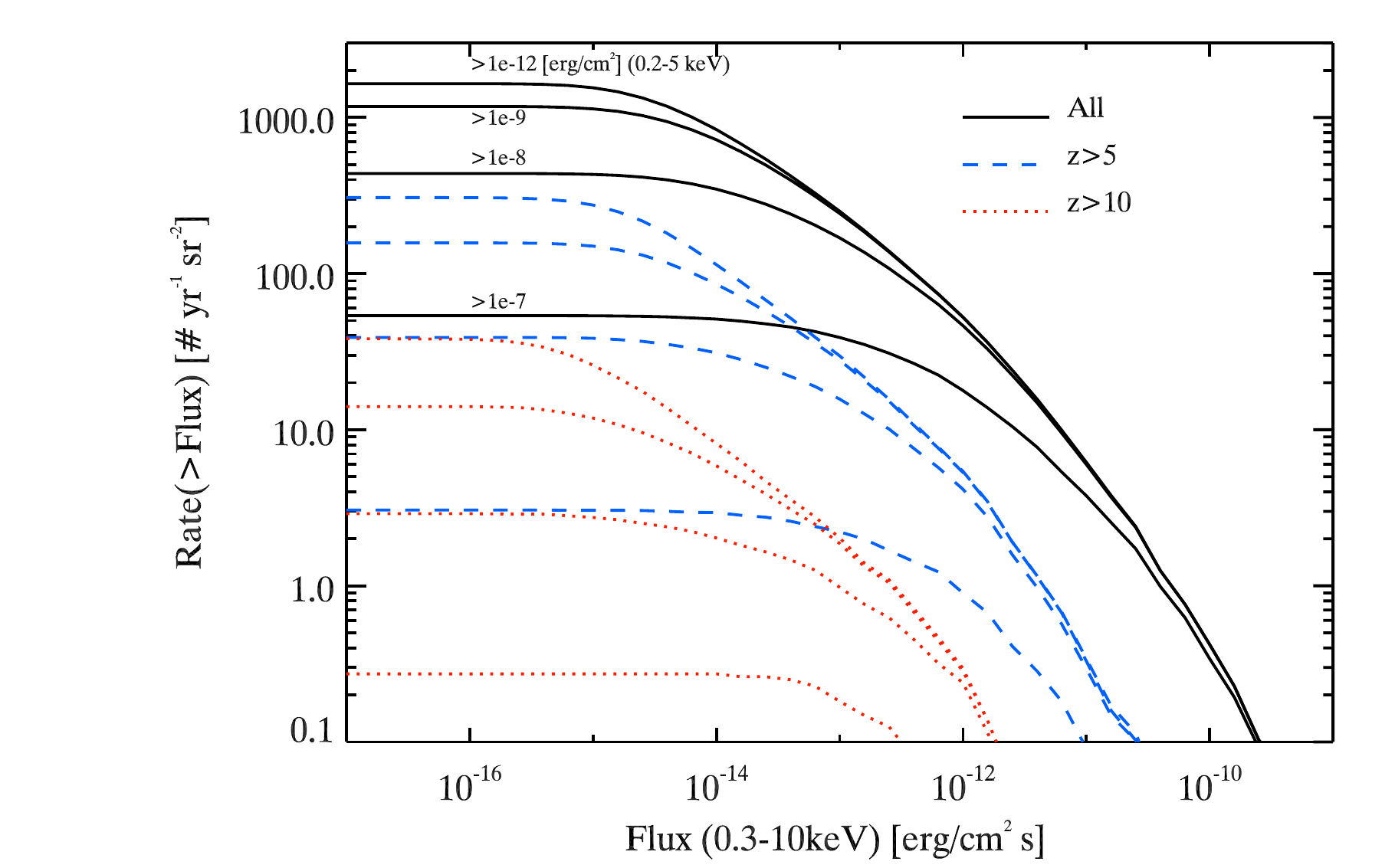}
\caption{
Cumulative distributions of the (0.3--10 keV) X-ray flux at 5 h after the burst. 
Same symbols as in Fig. \ref{fg5}.
}
\label{fg4a}
\end{figure}
 
Finally we derive the X--ray (0.3-10 keV) flux of the simulated GRB population. 
This is interesting for the use of GRBs as probes of the chemical composition and properties 
of the intervening medium (e.g. Campana et al. 2011) through the metal absorption 
edges imprinted on their X--ray spectra. 

The X--ray properties of the GRBs of the BAT6 sample have been presented in D'Avanzo et al. (2012). 
They found that the X--ray luminosity is correlated with both the prompt emission 
$\gamma$--ray isotropic luminosity and energy (\liso\ and \eiso\ respectively - Amati et al. 2002; Yonetoku et al. 2004). 
There is evidence that the early X--ray emission observed in GRBs has a different origin 
to the afterglow produced at external shocks (Ghisellini et al. 2009). Besides there could be 
a connection between the prompt and 
X--ray emission (Margutti et al., 2013). 
In particular, the early time emission in the X--ray band could be due to an emission 
component related to the long lived activity of the central engine (invoked also by 
late time X--ray flares observed in the \sw\ XRT light curves -- Burrows et al. 2007).  
Therefore, we estimated the X--ray flux of the simulated GRB population through 
the empirical correlations found by D'Avanzo et al. (2012).
Fig. \ref{fg4a} shows the cumulative distribution of the 
flux in the 0.3--10 keV energy range  for different cuts of the 0.2--5 keV fluence. 

With a Target Of Opportunity (TOO) reaction time of 4-5 hours Athena (Nandra et al. 2013) will be able to provide high resolution spectroscopy 
for a large number of GRB afterglows. In order to enable high resolution spectroscopy in the 
microcalorimeter (XIFU) instrument, that delivers 2.5 eV resolution, a total fluence of $10^{5}$ cts is needed (Jonker et al. 2013). 
This corresponds to  typical afterglow flux of $\ge10^{-12}$ erg cm$^{-2}$ s$^{-1}$ with a integration time of $\sim$50-100 ksec. With this flux limit, the expected rate of GRBs is 3(0.2) yr$^{-1}$ sr$^{-1}$ at $z\ge5$(10). 

The effective number of GRB that can be followed up by Athena wil depend on the field of view and the observational strategy of the GRB mission dedicated to detect GRBs and provide trigger alerts for the X--ray observations. For a limited ($\leq$1-3 sr) field of view of a soft X--ray detector (operated e.g. in the 0.2--5 keV), the optimal strategy to maximize the number of follow-up with Athena will be to point towards the sky accessible by Athena at that given time (taking into account its observing program). This strategy would  mimimize the slewing time in case of a trigger. In such a case the Athena observing efficiency will be close to 100\% and the TOO reaction time would be also minimized. For a much larger FOV, ($>$ 1/2 sky), the number of GRB follow up by Athena will, on the contrary reduce to ~50\%.



\section{Discussion}

We have presented the prompt and afterglow properties of a synthetic population of long GRBs. 
The main assumptions of the simulations are the luminosity function and the GRB formation rate 
(proportional to the star formation rate). 
GRB prompt emission properties are derived through the empirical correlations between 
the peak energy of their spectra and the isotropic equivalent energy and luminosity (\ama\ and \yone\ correlations). 
The simulated population is normalized to the bright sample of GRBs detected by 
\sw\ (i.e GRBs with peak flux $P_{\rm 15-150 keV}\ge 2.6$ ph cm$^{-2}$ s$^{-1}$). 
The simulated population also reproduces the properties (redshift distribution and $R$ afterglow 
flux distribution) of these real bursts. 
We have shown that the simulated population is representative also of the larger population 
of bursts detected by \fe\ and \ba. 

The detection rates presented are based on the best fit values of the assumed GRB formation rate (Eq. \ref{z}) and their luminosity function (Eq. \ref{lf}). The main source of uncertainty for the predictions of the rates of high redshift bursts is the strength $\delta$ of the evolution with redshift  of the GRB formation rate. Considering $\delta=1.7\pm0.5$ (S12), the rates can be larger/smaller by a factor $\sim$1.5. This only affects the absolute rate of detection  and not the comparison of the fraction of bursts accessible with instruments working in different energy bands as those presented here (\S 4.2, 4.3).

In general our results show that the largest rate of high--$z$ GRBs can be detected with 
a soft X--ray detector (e.g. operated in the 0.2--5 keV energy range like a Lobster-type wide-field focussing X-ray) because this is 
where the bulk of the high--$z$ GRBs have their peak energy (due to the cosmological redshift). 
However, such an instrument should be designed so to have a sensitivity that accounts properly 
for the the presence of a link between the \ep\ and the luminosity/energy of GRBs (\yone\ and  \ama\ correlations). 
To asses a population of GRBs say a factor of ten softer in \ep\, requires a sensitivity improvement 
of a factor $\sim$100. This is due to the slope of the observer frame distribution of the points in Fig.\ref{fg6} which is 
a result of the \yone\ correlation.  
The purpose of this work is to provide a tool for designing instruments for 
detecting a large sample of GRBs to study the high--$z$ Universe. 
 
The results presented so far are based on the set of assumptions described in \S2: the evolution of the GRB formation rate with redshift (Eq.~1) and the \ama\ and \yone\ correlations which are used, considering also their scatter, to link the energetic/luminosity of the simulated bursts with their peak energy. 

The first assumption is motivated by the fact that an evolution of the GRB density is also consistent with the metallicity bias as found from the study of GRB hosts (e.g. Vergani et al. 2014), and it is theoretically expected within the collapsar model. However, also the alternative scenario in which the luminosity function  evolves with redshift has been shown to be consistent with the observations (e.g. Salvaterra et al. 2012). For completeness we have also explored this possibility, assuming the GRB rate without any redshift evolution (i.e. $\delta=0$ in Eq.1) but, instead, a luminosity function described by a broken powerlaw with a break luminosity $L_{\rm cut}\propto(1+z)^{2.1}$. This model increases the fraction of luminous bursts at high redshifts: the results are shown in Fig.\ref{fgA2} by the dashed lines. Compared with the results obtained with an evolving GRB rate (\S2 - shown by the solid lines in Fig.\ref{fgA2}), the  global effect is to reduce the rate of dim bursts due to the increase of the high redshift and bright bursts (bottom panel of Fig.\ref{fgA2}). However, these differences are consistent with the typical uncertainty on the estimated rates (shown by the vertical bar in Fig.\ref{fgA2}). 

The other assumptions,  i.e. the \ama\ and \yone\ correlations can also be tested. We have assumed (\S2) these two correlations with their scatter. However, it is still debated if they are due to instrumental selection effects which, if present, should limit the detection of GRBs of intermediate/low luminosity/energy, i.e. those on the left hand side of these correlations (in the \ama\ and \yone\ correlations we typically plot the x--axis representing the isotropic energy \eiso\ or luminosity \liso).  Although excluded by recent simulations (Ghirlanda et al. 2012) 
we have considered the extreme case of a uniform distribution of simulated bursts on the left hand side of the \ama\ and \yone\ correlations. 
In practice we assume that the \yone\ and \ama\ correlations are boundaries in their respective planes (e.g. Nakar \& Piran 2005; Butler Bloom \& Poznanski 2010) which are, therefore divided in two regions: on the right hand side of the correlations there are no bursts (if they were present we should have detected) while on the left hand side there is no intrinsic correlation between e.g. \ep\ and \liso(\eiso).  The results obtained with this  assumption are shown by the dashed lines in Fig.\ref{fgA3}. From the comparison with the results (solid lines) obtained with the set of assumptions of \S2 (i.e. assuming instead the correlations with their scatter), it appears that the difference is minimal and consistent within the uncertainty on the estimated rates.

\section{Conclusion}

Through our code we can derive the prompt emission flux and fluence and the afterglow flux 
(at any time and frequency) of the population of GRBs distributed up to $z=20$. 
We note that our code does not include the possibility that the GRB rate at high 
redshift is enhanced by the primordial massive (popIII) stars which could even be dominating
at $z\ge10$ (Campisi et al. 2011). 
We have derived for the first time also the afterglow $J$ and $H$ band flux for high redshift GRBs. 
This is important because it can be used to assess the fraction of detected GRBs whose afterglow 
can be observed (or spectrum acquired) with an IR telescope (in order to maximize the study of the high--$z$ population). 

The scope of this paper is to provide a tool to estimate the detection rate of GRBs and, 
in particular, of high--$z$ bursts. 
High--$z$ GRBs are of particular relevance for planning forthcoming dedicated missions 
to study the high--$z$ Universe. 
We have shown here the flux and fluence rate distributions considering four possible 
energy bands where present and future GRB detectors can operate. 
The presented curves can be used to estimate approximately the detection rates expected 
for an instrument operating in a given energy range  and with a given sensitivity threshold.  

\section*{Acknowledgments}
We acknowledge a Prin--INAF 2011 grant (1.05.01.09.15) and ASI--INAF contract (I/004/11/1). 
We thank the anonymous referee for his/her valuable comments that improved the manuscript. 
Specific simulations for energy ranges not considered in this paper can be provided on request.  
JPO acknowledges partial support from the UK Space Agency. DB is funded through ARC grant DP110102034.
DG acknowledges the financial support of the UnivEarthS Labex program at
Sorbonne Paris Cit\'e (ANR-10-LABX-0023 and ANR-11-IDEX-0005-02)

\newpage
\clearpage
\section*{Appendix}
\begin{figure}
\includegraphics[scale=0.45,trim=0.2cm 13cm 0.2cm 3.5cm, clip]{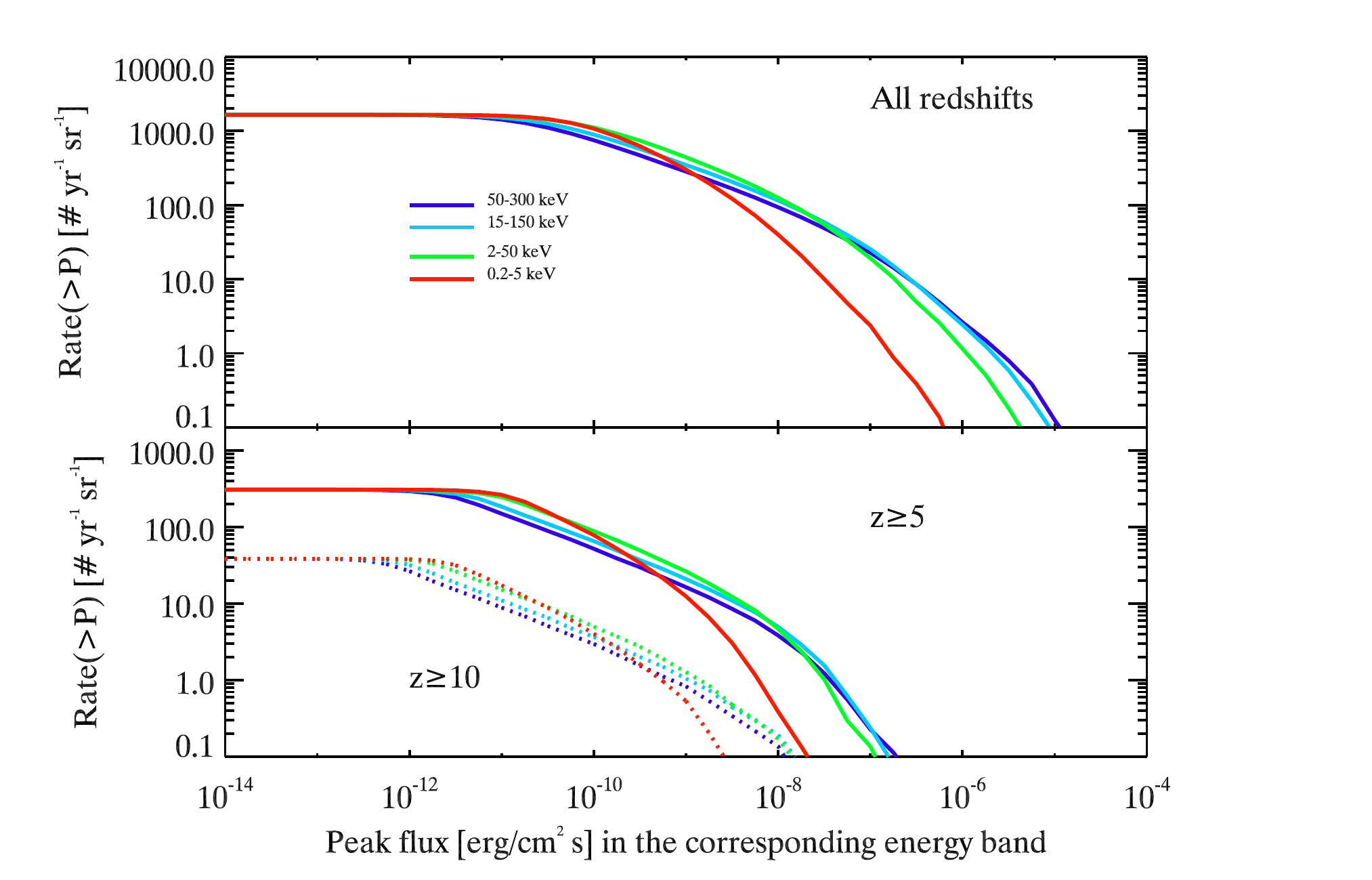}
\caption{Cumulative distributions of the peak flux (in units of erg cm$^{-2}$ s$^{-1}$). Same lines and color coding of Fig.\ref{fg3}. }
\label{fgA0}
\end{figure}

\begin{figure}
\includegraphics[scale=0.45,trim=0.2cm 13cm 0.2cm 3.5cm, clip]{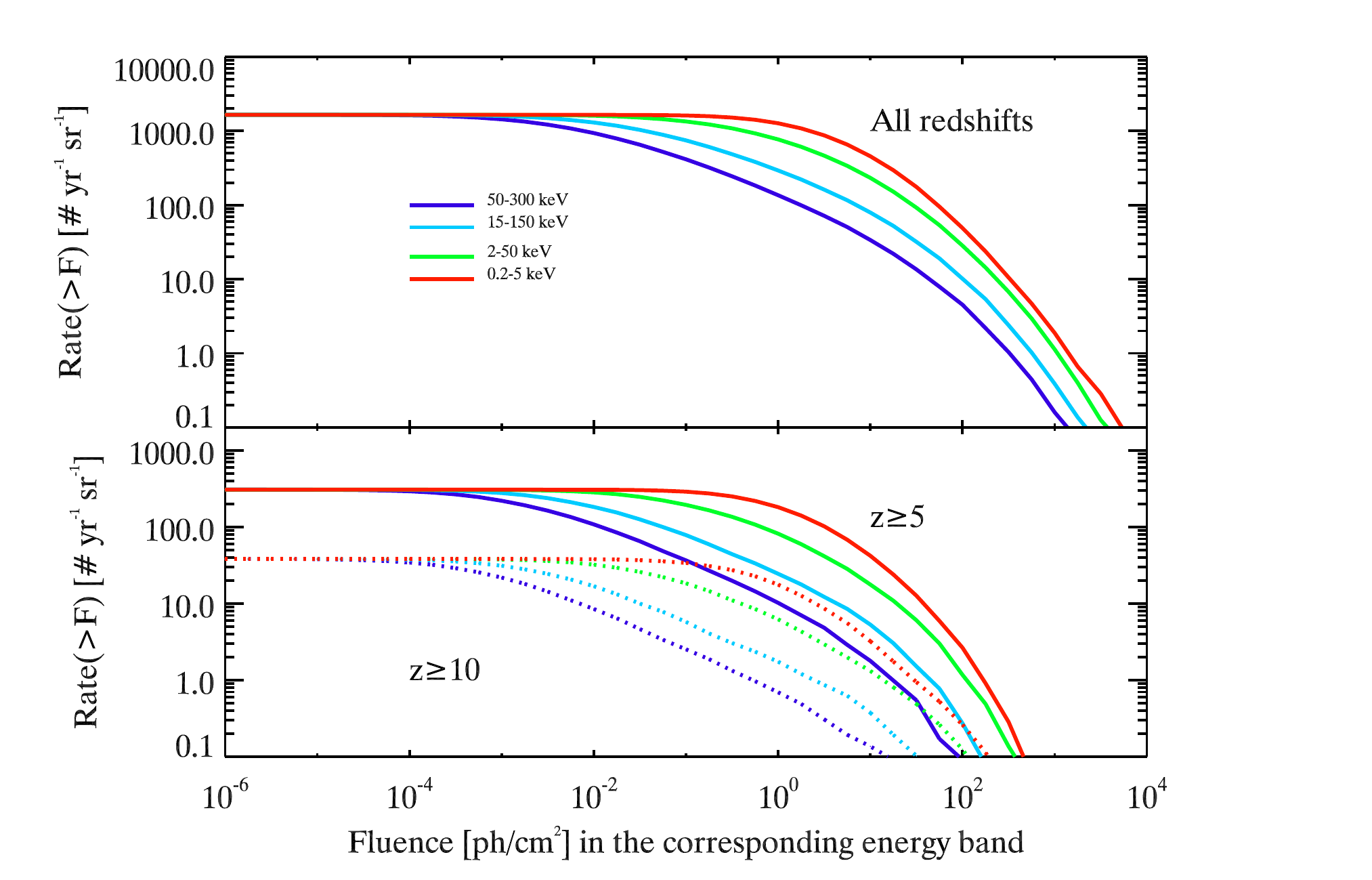}
\caption{Cumulative distributions of the fluence (in units of ph cm$^{-2}$). Same lines and color coding of Fig.\ref{fg4}.}
\label{fgA1}
\end{figure}

\begin{figure}
\includegraphics[scale=0.45,trim=0.2cm 13cm 0.2cm 3.5cm, clip]{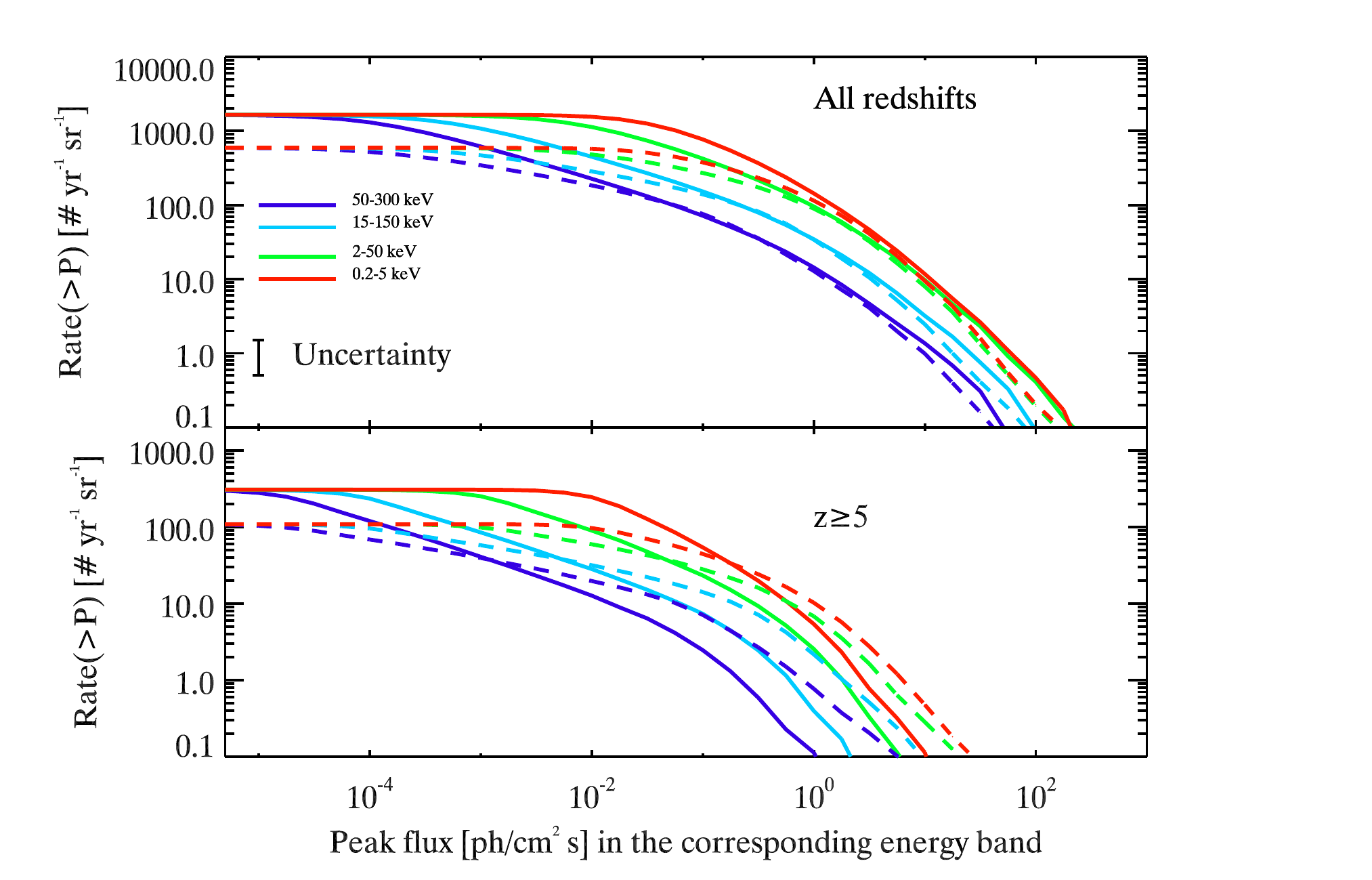}
\caption{ Cumulative distributions of the peak flux  comparing the results obtained assuming the evolution of the GRB formation rate (solid lines - \S2) and the evolution of the luminosity function (dashed lines). Same lines and color coding of Fig.\ref{fg3}. 
}
\label{fgA2}
\end{figure}

We show in Fig.\ref{fgA0}  the cumulative distributions of the peak flux in units of erg cm$^{-2}$ s$^{-1}$ and in Fig.\ref{fgA1} of the fluence in units of ph cm$^{-2}$. 
In the results presented in the paper we have assumed a density evolution of the GRB formation rate with a constant luminosity function, i.e. a broken powerlaw (\S2). If the evolution is instead assigned to the luminosity function, we adopt the model parameters of Salvaterra et al. (2012) where there is an evolution of the cut of the luminosity function $L_{\rm cut}\propto (1+z)^{2.1}$. This evolution increases the number of high luminosity GRBs at high redshifts. Fig.\ref{fgA2} compares the cumulative peak flux distribution obtained with the assumption of a density evolution of GRBs (\S2 - solid lines) and with the assumption of a luminosity evolution (dashed lines in Fig.\ref{fgA2}). 

\begin{figure}
\includegraphics[scale=0.45,trim=0.2cm 13cm 0.2cm 3.5cm, clip]{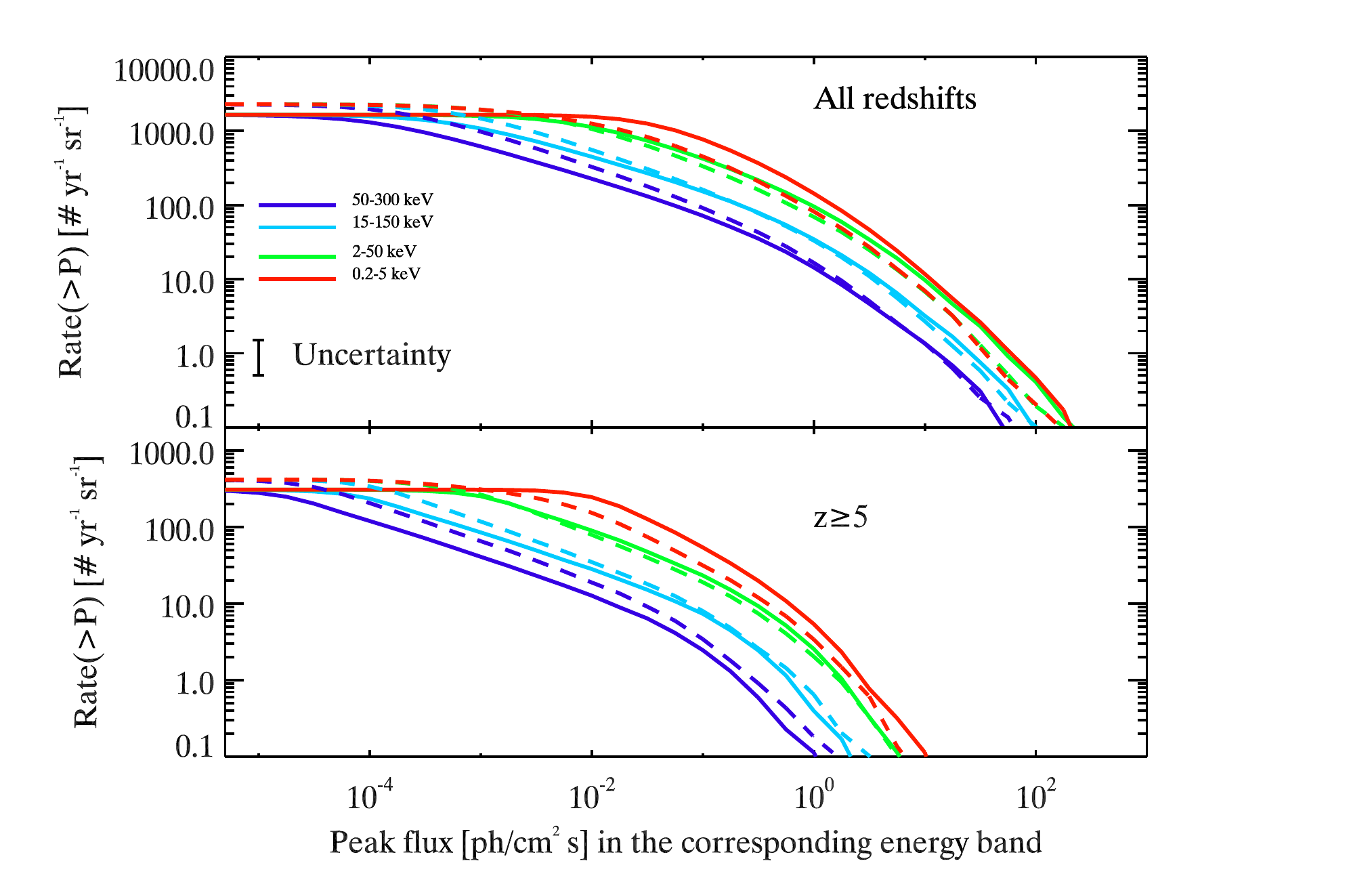}
\caption{ Cumulative distributions of the peak flux comparing the results obtained with the assumptions of \S2 (solid lines) and that the \ama\ and \yone\ correlations are boundaries in the corresponding planes (dashed lines).  Same lines and color coding of Fig.\ref{fg3}. 
}
\label{fgA3}
\end{figure}

The other main hypothesis of our paper is the existence of the \ama\ and \yone\ correlations. It has been discussed  that these could boundaries in the corresponding planes and some instrumental/observational effects prevent the exploration of the part of the planes which would be uniformly filled with intermediate/low \eiso (\liso) bursts. We have tested this possibility with our code and show in Fig.\ref{fgA3} our results. We have assumed the \ama\ and \yone\ correlations, as obtained from the current data, and assumed that they divide the plane into two regions: the right hand side of these correlation is free of bursts, since if any GRB would be in this semi--plane it should have been detected, while the left hand side of the correlation is filled (uniformly) with bursts which have their peak energy uncorrelated with the \eiso\ or \liso. The boundary of the uniformly filled semi--plane are defined assuming that bursts have \ep$\in(0.01;10^{4})$ keV. This different assumption produce a flux cumulative distribution (dashed line in Fig.\ref{fgA3}) which is only slightly different from that obtained assuming the correlations as they are observed with their actual scatter (solid lines in Fig.\ref{fgA3}). 


\end{document}